
%
%
\newif\iffigmode
\figmodetrue
\iffigmode\input epsf.tex\else\message{THIS WILL NOT INCLUDE THE
FIGURE}\fi
%
\expandafter \def \csname CHAPLABELintro\endcsname {1}
\expandafter \def \csname EQLABELgenyukawa\endcsname {1.1?}
\expandafter \def \csname EQLABELgenexpansion\endcsname {1.2?}
\expandafter \def \csname CHAPLABELclassical\endcsname {2}
\expandafter \def \csname EQLABELvident\endcsname {2.1?}
\expandafter \def \csname EQLABELTdef\endcsname {2.2?}
\expandafter \def \csname EQLABELdefm\endcsname {2.3?}
\expandafter \def \csname EQLABELpolend\endcsname {2.4?}
\expandafter \def \csname EQLABELcicyR\endcsname {2.5?}
\expandafter \def \csname EQLABELDdef\endcsname {2.6?}
\expandafter \def \csname EQLABELsexp\endcsname {2.7?}
\expandafter \def \csname EQLABELsseq\endcsname {2.8?}
\expandafter \def \csname EQLABELsss\endcsname {2.9?}
\expandafter \def \csname CHAPLABELgeneral\endcsname {3}
\expandafter \def \csname EQLABELsaction\endcsname {3.1?}
\expandafter \def \csname EQLABELbinstanton\endcsname {3.2?}
\expandafter \def \csname EQLABELfinstantons\endcsname {3.3?}
\expandafter \def \csname FIGLABELfinst\endcsname {3.1?}
\expandafter \def \csname EQLABELinst.eq\endcsname {3.4?}
\expandafter \def \csname EQLABELinstantini\endcsname {3.5?}
\expandafter \def \csname EQLABELV10\endcsname {3.6?}
\expandafter \def \csname EQLABELV0\endcsname {3.7?}
\expandafter \def \csname EQLABELV10*\endcsname {3.8?}
\expandafter \def \csname EQLABELV0*\endcsname {3.9?}
\expandafter \def \csname EQLABELodef\endcsname {3.10?}
\expandafter \def \csname EQLABELV1\endcsname {3.11?}
\expandafter \def \csname EQLABELVR\endcsname {3.12?}
\expandafter \def \csname EQLABELdecomp\endcsname {3.13?}
\expandafter \def \csname EQLABELpotentials\endcsname {3.14?}
\expandafter \def \csname EQLABELintegrals\endcsname {3.15?}
\expandafter \def \csname CHAPLABELmass\endcsname {4}
\expandafter \def \csname EQLABELRSS\endcsname {4.1?}
\expandafter \def \csname EQLABELdeterminants\endcsname {4.2?}
\expandafter \def \csname EQLABELQT\endcsname {4.3?}
\expandafter \def \csname EQLABELmasses\endcsname {4.4?}
\expandafter \def \csname CHAPLABELquintic\endcsname {5}
\expandafter \def \csname EQLABELmdef\endcsname {5.1?}
\expandafter \def \csname EQLABELvecdef\endcsname {5.2?}
\expandafter \def \csname EQLABELsingrep\endcsname {5.3?}
\expandafter \def \csname EQLABELtwolines\endcsname {5.4?}
\expandafter \def \csname EQLABELpeps\endcsname {5.5?}
\expandafter \def \csname EQLABELsings\endcsname {5.6?}
\expandafter \def \csname EQLABELminusoneminusone\endcsname {5.7?}
\expandafter \def \csname EQLABELinstmetric\endcsname {5.8?}
\expandafter \def \csname EQLABELmpole\endcsname {5.9?}
\expandafter \def \csname EQLABELthirdline\endcsname {5.10?}
\expandafter \def \csname EQLABELpeps2\endcsname {5.11?}
\expandafter \def \csname EQLABELminusoneminusone2\endcsname {5.12?}
\expandafter \def \csname EQLABELnodalP\endcsname {5.13?}
\expandafter \def \csname EQLABELmirrorpoly\endcsname {5.14?}
\expandafter \def \csname EQLABELXsharp\endcsname {5.15?}
\expandafter \def \csname CHAPLABELyuk\endcsname {6}
\expandafter \def \csname EQLABELWhew!\endcsname {6.1?}
\expandafter \def \csname EQLABELWhew.\endcsname {6.2?}
\expandafter \def \csname EQLABELderivative\endcsname {6.3?}
\expandafter \def \csname EQLABELsinglets1\endcsname {6.4?}
\expandafter \def \csname EQLABELCauchy\endcsname {6.5?}
\expandafter \def \csname EQLABELsinglets2\endcsname {6.6?}
\expandafter \def \csname CHAPLABELconc\endcsname {7}
\expandafter \def \csname CHAPLABELappendix\endcsname {-2}
\expandafter \def \csname EQLABELzmodebasis\endcsname {-2.1?}
\expandafter \def \csname EQLABELmetrics\endcsname {-2.2?}
\expandafter \def \csname EQLABELdetprime\endcsname {-2.3?}
\expandafter \def \csname EQLABELmovingbasis\endcsname {-2.4?}
\expandafter \def \csname EQLABELcoordchange\endcsname {-2.5?}
\expandafter \def \csname EQLABELlzeromodes\endcsname {-2.6?}
%
%
%

\font\eightrm=cmr8 at 8pt
\font\tenrm=cmr12 at 10pt

\font\seventeenrm=cmr17 at 17pt
\font\twentyonerm=cmr17 at 21pt

\font\ss=cmss10

\font\csc=cmcsc10

\font\twelvecal=cmsy10 at 12pt

\font\twelvemath=cmmi12
\font\fourteenmath=cmmi12 at 14pt
\font\eightboldmath=cmmib10 at 8pt

\font\seventeenbold=cmbx7 at 17pt

\font\fively=lasy5
\font\sevenly=lasy7
\font\tenly=lasy10

\textfont10=\tenly
\scriptfont10=\sevenly
\scriptscriptfont10=\fively
\magnification=1200
\parskip=10pt
\parindent=20pt
\def\today{\ifcase\month\or January\or February\or March\or April\or May\or
June
       \or July\or August\or September\or October\or November\or December\fi
       \space\number\day, \number\year}

\def\title#1{\footline={\ifnum\pageno<2\hfil
       \else\hss\tenrm\folio\hss\fi}\vskip1truein\centerline{{#1}
       \footnote{\raise1ex\hbox{*}}{\eightrm Supported in part
       by the Robert A. Welch Foundation and N.S.F. Grants
       PHY-880637 and\break PHY-8605978.}}}

\def\Z{\hfill\break}
\def\newpage{\vfill\eject}
\def\abstract#1{\centerline{\bf ABSTRACT}\vskip.2truein{\narrower\noindent#1
       \smallskip}}

\def\runninghead#1#2{\voffset=2\baselineskip\nopagenumbers
       \headline={\ifodd\pageno\rightheadline\else \leftheadline\fi}
       \def\rightheadline{{\sl#1}\hfill{\rm\folio}}
       \def\leftheadline{{\rm\folio}\hfill{\sl#2}}}
\def\SS{\mathhexbox278}

\newcount\footnoteno
\def\Footnote#1{\advance\footnoteno by 1
                \let\SF=\empty
                \ifhmode\edef\SF{\spacefactor=\the\spacefactor}\/\fi
                $^{\the\footnoteno}$\ignorespaces
                \SF\vfootnote{$^{\the\footnoteno}$}{#1}}

\def\figbox#1#2#3{\vbox{\vskip15pt
                   \vbox{\hrule
                    \hbox{\vrule
                     \vbox{\vskip12truept\centerline #1 \vskip6truept
                          {\hskip.4truein\vbox{\hsize=5truein\noindent
                          {\bf Figure\hskip5truept#2:}\hskip5truept#3}}
                     \vskip18truept}
                    \vrule}
                   \hrule}}}
\def\place#1#2#3{\vbox to0pt{\kern-\parskip\kern-7pt
                             \kern-#2truein\hbox{\kern#1truein #3}
                             \vss}\nointerlineskip}
\def\figurecaption#1#2{\kern.75truein\vbox{\hsize=5truein\noindent{\bf Figure
    \figlabel{#1}:} #2}}
\def\tablecaption#1#2{\kern.75truein\lower12truept\hbox{\vbox{\hsize=5truein
    \noindent{\bf Table\hskip5truept\tablabel{#1}:} #2}}}
\def\boxed#1{\lower3pt\hbox{
                       \vbox{\hrule\hbox{\vrule

\vbox{\kern2pt\hbox{\kern3pt#1\kern3pt}\kern3pt}\vrule}
                         \hrule}}}
\def\a{\alpha}
\def\b{\beta}
\def\g{\gamma}\def\G{\Gamma}
\def\d{\delta}
\def\e{\epsilon}\def\ve{\varepsilon}
\def\z{\zeta}

\def\vth{\vartheta}

\def\k{\kappa}
\def\l{\lambda}
\def\m{\mu}
\def\n{\nu}
\def\x{\xi}

\def\p{\pi}
\def\r{\rho}
\def\s{\sigma}\def\S{\Sigma}
\def\t{\tau}

\def\ph{\phi}\def\vph{\varphi}
\def\ch{\chi}
\def\ps{\psi}
\def\o{\omega}\def\O{\Omega}

\def\ca#1{\relax\ifmmode {{\cal #1}}\else $\cal #1$\fi}

\def\calb{{\cal B}}

\def\calm{{\cal M}}

\def\inbar{\vrule height1.5ex width.4pt depth0pt}
\def\IB{\relax{\rm I\kern-.18em B}}
\def\IC{\relax\hbox{\kern.25em$\inbar\kern-.3em{\rm C}$}}
\def\ID{\relax{\rm I\kern-.18em D}}
\def\IE{\relax{\rm I\kern-.18em E}}
\def\IF{\relax{\rm I\kern-.18em F}}
\def\IG{\relax\hbox{\kern.25em$\inbar\kern-.3em{\rm G}$}}
\def\IH{\relax{\rm I\kern-.18em H}}
\def\II{\relax{\rm I\kern-.18em I}}
\def\IK{\relax{\rm I\kern-.18em K}}
\def\IL{\relax{\rm I\kern-.18em L}}
\def\IM{\relax{\rm I\kern-.18em M}}
\def\IN{\relax{\rm I\kern-.18em N}}
\def\IO{\relax\hbox{\kern.25em$\inbar\kern-.3em{\rm O}$}}
\def\IP{\relax{\rm I\kern-.18em P}}
\def\IQ{\relax\hbox{\kern.25em$\inbar\kern-.3em{\rm Q}$}}
\def\IR{\relax{\rm I\kern-.18em R}}
\def\IZ{\relax\ifmmode\hbox{\ss Z\kern-.4em Z}\else{\ss Z\kern-.4em Z}\fi}
\def\IGa{\relax{\rm I}\kern-.18em\Gamma}
\def\IPi{\relax{\rm I}\kern-.18em\Pi}
\def\ITh{\relax\hbox{\kern.25em$\inbar\kern-.3em\Theta$}}
\def\IOm{\relax\thinspace\inbar\kern1.95pt\inbar\kern-5.525pt\Omega}


\def\ie{{\it i.e.\ \/}}

\def\noblackboxes{\overfullrule=0pt}
\def\define{\buildrel\rm def\over =}

\def\cy{Calabi--Yau}
\def\cym{Calabi--Yau manifold}

\def\K{K\"ahler}

\def\H#1#2{\relax\ifmmode {H^{#1#2}}\else $H^{#1 #2}$\fi}
\def\M{\relax\ifmmode{\calm}\else $\calm$\fi}

\def\Bigcheck{\lower3.8pt\hbox{\smash{\hbox{{\twentyonerm \v{}}}}}}
\def\bigboldcheck{\smash{\hbox{{\seventeenbold\v{}}}}}

\def\Bighat{\lower3.8pt\hbox{\smash{\hbox{{\twentyonerm \^{}}}}}}

\def\Msharp{\relax\ifmmode{\calm^\sharp}\else $\smash{\calm^\sharp}$\fi}
\def\Mflat{\relax\ifmmode{\calm^\flat}\else $\smash{\calm^\flat}$\fi}
\def\preMcheck{\kern2pt\hbox{\Bigcheck\kern-12pt{$\cal M$}}}
\def\Mcheck{\relax\ifmmode\preMcheck\else $\preMcheck$\fi}
\def\preMhat{\kern2pt\hbox{\Bighat\kern-12pt{$\cal M$}}}
\def\Mhat{\relax\ifmmode\preMhat\else $\preMhat$\fi}

\def\Bsharp{\relax\ifmmode{\calb^\sharp}\else $\calb^\sharp$\fi}
\def\Bflat{\relax\ifmmode{\calb^\flat}\else $\calb^\flat$ \fi}
\def\preBcheck{\hbox{\Bigcheck\kern-9pt{$\cal B$}}}
\def\Bcheck{\relax\ifmmode\preBcheck\else $\preBcheck$\fi}
\def\preBhat{\hbox{\Bighat\kern-9pt{$\cal B$}}}
\def\Bhat{\relax\ifmmode\preBhat\else $\preBhat$\fi}

\def\figBcheck{\kern3pt\hbox{\raise1pt\hbox{\bigboldcheck}\kern-11pt
    {\twelvecal B}}}
\def\figBsharp{{\twelvecal B}\raise5pt\hbox{$\twelvemath\sharp$}}
\def\figBflat{{\twelvecal B}\raise5pt\hbox{$\twelvemath\flat$}}

\def\gcheck{\hbox{\lower2.5pt\hbox{\Bigcheck}\kern-8pt$\g$}}
\def\lhat{\hbox{\raise.5pt\hbox{\Bighat}\kern-8pt$\l$}}

\def\Fcheck{\kern2pt\hbox{\raise1pt\hbox{\Bigcheck}\kern-10pt{$\cal F$}}}
\def\Fhat{\kern2pt\hbox{\raise1pt\hbox{\Bighat}\kern-10pt{$\cal F$}}}

\def\boldIP{\relax{\tenboldrm \char '111 \kern-.18em \char '120}}
\def\boldcp#1{\relax{\boldIP\kern-2pt\lower.5ex\hbox{\eightboldmath #1}}}
\def\cp#1{\relax\ifmmode {\IP\kern-2pt{}_{#1}}\else $\IP\kern-2pt{}_{#1}$\fi}
\def\h#1#2{\relax\ifmmode {b_{#1#2}}\else $b_{#1#2}$\fi}
\def\Z{\hfill\break}
\def\imag{\Im m}
\def\half{{1\over 2}}

\def\frac#1#2{{#1\over #2}}

\def\pd#1#2{{\partial #1\over\partial #2}}
\def\ppd#1#2#3{{\partial^2 #1\over \partial #2\partial #3}}

\def\ex#1{{\bf Exercise:}~~#1\hfill$\diamondsuit$}
\def\cone{\relax\thinspace\hbox{$<\kern-.8em{)}$}}
\mathchardef\mho"0A30

\def\asymp{\sim}
\def\-{\hphantom{-}}


\def\npb#1{Nucl.\ Phys.\ {\bf B#1}}

\def\cmp#1{Commun. Math. Phys. {\bf #1}}
\def\plb#1{Phys. Lett. {\bf #1B}}


\def\picture #1 by #2 (#3){\vbox to #2{\hrule width #1 height 0pt depth 0pt
                                       \vfill\special{picture #3}}}
\def\scaledpicture #1 by #2 (#3 scaled #4){{\dimen0=#1 \dimen1=#2
           \divide\dimen0 by 1000 \multiply\dimen0 by #4
            \divide\dimen1 by 1000 \multiply\dimen1 by #4
            \picture \dimen0 by \dimen1 (#3 scaled #4)}}
\def\illustration #1 by #2 (#3){\vbox to #2{\hrule width #1 height 0pt depth
0pt
                                       \vfill\special{illustration #3}}}
\def\scaledillustration #1 by #2 (#3 scaled #4){{\dimen0=#1 \dimen1=#2
           \divide\dimen0 by 1000 \multiply\dimen0 by #4
            \divide\dimen1 by 1000 \multiply\dimen1 by #4
            \illustration \dimen0 by \dimen1 (#3 scaled #4)}}


\def\delaOssa{\nobreak\vskip1truein\hbox to\hsize
       {\hskip 4truein Xenia de la Ossa\hfill}}

\def\hoy{\number\day\space de \ifcase\month\or enero\or febrero\or marzo\or
       abril\or mayo\or junio\or julio\or agosto\or septiembre\or octubre\or
       noviembre\or diciembre\fi\space de \number\year}


\newif\ifproofmode
\proofmodefalse

\newif\ifforwardreference
\forwardreferencefalse

\newif\ifchapternumbers
\chapternumbersfalse

\newif\ifcontinuousnumbering
\continuousnumberingfalse

\newif\iffigurechapternumbers
\figurechapternumbersfalse

\newif\ifcontinuousfigurenumbering
\continuousfigurenumberingfalse

\newif\iftablechapternumbers
\tablechapternumbersfalse

\newif\ifcontinuoustablenumbering
\continuoustablenumberingfalse

\font\eqsixrm=cmr6

\def\marginstyle{\eqsixrm}

\newtoks\chapletter
\newcount\chapno
\newcount\eqlabelno
\newcount\figureno
\newcount\tableno

\chapno=0
\eqlabelno=0
\figureno=0
\tableno=0

\def\chapfolio{\ifnum\chapno>0 \the\chapno\else\the\chapletter\fi}

\def\bumpchapno{\ifnum\chapno>-1 \global\advance\chapno by 1
\else\global\advance\chapno by -1 \setletter\chapno\fi
\ifcontinuousnumbering\else\global\eqlabelno=0 \fi
\ifcontinuousfigurenumbering\else\global\figureno=0 \fi
\ifcontinuoustablenumbering\else\global\tableno=0 \fi}

\def\setletter#1{\ifcase-#1{}\or{}%
\or\global\chapletter={A}%
\or\global\chapletter={B}%
\or\global\chapletter={C}%
\or\global\chapletter={D}%
\or\global\chapletter={E}%
\or\global\chapletter={F}%
\or\global\chapletter={G}%
\or\global\chapletter={H}%
\or\global\chapletter={I}%
\or\global\chapletter={J}%
\or\global\chapletter={K}%
\or\global\chapletter={L}%
\or\global\chapletter={M}%
\or\global\chapletter={N}%
\or\global\chapletter={O}%
\or\global\chapletter={P}%
\or\global\chapletter={Q}%
\or\global\chapletter={R}%
\or\global\chapletter={S}%
\or\global\chapletter={T}%
\or\global\chapletter={U}%
\or\global\chapletter={V}%
\or\global\chapletter={W}%
\or\global\chapletter={X}%
\or\global\chapletter={Y}%
\or\global\chapletter={Z}\fi}

\def\tempsetletter#1{\ifcase-#1{}\or{}%
\or\global\chapletter={A}%
\or\global\chapletter={B}%
\or\global\chapletter={C}%
\or\global\chapletter={D}%
\or\global\chapletter={E}%
\or\global\chapletter={F}%
\or\global\chapletter={G}%
\or\global\chapletter={H}%
\or\global\chapletter={I}%
\or\global\chapletter={J}%
\or\global\chapletter={K}%
\or\global\chapletter={L}%
\or\global\chapletter={M}%
\or\global\chapletter={N}%
\or\global\chapletter={O}%
\or\global\chapletter={P}%
\or\global\chapletter={Q}%
\or\global\chapletter={R}%
\or\global\chapletter={S}%
\or\global\chapletter={T}%
\or\global\chapletter={U}%
\or\global\chapletter={V}%
\or\global\chapletter={W}%
\or\global\chapletter={X}%
\or\global\chapletter={Y}%
\or\global\chapletter={Z}\fi}

\def\chapshow#1{\ifnum#1>0 \relax#1%
\else{\tempsetletter{\number#1}\chapno=#1\chapfolio}\fi}

\def\chaplabel#1{\bumpchapno\ifproofmode\ifforwardreference
\immediate\write1{\noexpand\expandafter\noexpand\def
\noexpand\csname CHAPLABEL#1\endcsname{\the\chapno}}\fi\fi
\global\expandafter\edef\csname CHAPLABEL#1\endcsname
{\the\chapno}\ifproofmode\llap{\hbox{\marginstyle #1\ }}\fi\chapfolio}

\def\chapref#1{\ifundefined{CHAPLABEL#1}??\ifproofmode\ifforwardreference%
\else\write16{ ***Undefined Chapter Reference #1*** }\fi
\else\write16{ ***Undefined Chapter Reference #1*** }\fi
\else\edef\LABxx{\getlabel{CHAPLABEL#1}}\chapshow\LABxx\fi
\ifproofmode\write2{Chapter #1}\fi}

\def\eqnum{\global\advance\eqlabelno by 1
\eqno(\ifchapternumbers\chapfolio.\fi\the\eqlabelno)}

\def\eqlabel#1{\global\advance\eqlabelno by 1 \ifproofmode\ifforwardreference
\immediate\write1{\noexpand\expandafter\noexpand\def
\noexpand\csname EQLABEL#1\endcsname{\the\chapno.\the\eqlabelno?}}\fi\fi
\global\expandafter\edef\csname EQLABEL#1\endcsname
{\the\chapno.\the\eqlabelno?}\eqno(\ifchapternumbers\chapfolio.\fi
\the\eqlabelno)\ifproofmode\rlap{\hbox{\marginstyle #1}}\fi}

\def\eqalignnum{\global\advance\eqlabelno by 1
&(\ifchapternumbers\chapfolio.\fi\the\eqlabelno)}

\def\eqalignlabel#1{\global\advance\eqlabelno by 1 \ifproofmode
\ifforwardreference\immediate\write1{\noexpand\expandafter\noexpand\def
\noexpand\csname EQLABEL#1\endcsname{\the\chapno.\the\eqlabelno?}}\fi\fi
\global\expandafter\edef\csname EQLABEL#1\endcsname
{\the\chapno.\the\eqlabelno?}&(\ifchapternumbers\chapfolio.\fi
\the\eqlabelno)\ifproofmode\rlap{\hbox{\marginstyle #1}}\fi}

\def\eqref#1{\hbox{(\ifundefined{EQLABEL#1}***)\ifproofmode\ifforwardreference%
\else\write16{ ***Undefined Equation Reference #1*** }\fi
\else\write16{ ***Undefined Equation Reference #1*** }\fi
\else\edef\LABxx{\getlabel{EQLABEL#1}}%
\def\LAByy{\expandafter\stripchap\LABxx}\ifchapternumbers%
\chapshow{\LAByy}.\expandafter\stripeq\LABxx%
\else\ifnum\number\LAByy=\chapno\relax\expandafter\stripeq\LABxx%
\else\chapshow{\LAByy}.\expandafter\stripeq\LABxx\fi\fi)\fi}%
\ifproofmode\write2{Equation #1}\fi}

\def\fignum{\global\advance\figureno by 1
\relax\iffigurechapternumbers\chapfolio.\fi\the\figureno}

\def\figlabel#1{\global\advance\figureno by 1
\relax\ifproofmode\ifforwardreference
\immediate\write1{\noexpand\expandafter\noexpand\def
\noexpand\csname FIGLABEL#1\endcsname{\the\chapno.\the\figureno?}}\fi\fi
\global\expandafter\edef\csname FIGLABEL#1\endcsname
{\the\chapno.\the\figureno?}\iffigurechapternumbers\chapfolio.\fi
\ifproofmode\llap{\hbox{\marginstyle#1
\kern1.2truein}}\relax\fi\the\figureno}

\def\figref#1{\hbox{\ifundefined{FIGLABEL#1}!!!!
\ifproofmode\ifforwardreference%
\else\write16{ ***Undefined Figure Reference #1*** }\fi
\else\write16{ ***Undefined Figure Reference #1*** }\fi
\else\edef\LABxx{\getlabel{FIGLABEL#1}}%
\def\LAByy{\expandafter\stripchap\LABxx}\iffigurechapternumbers%
\chapshow{\LAByy}.\expandafter\stripeq\LABxx%
\else\ifnum \number\LAByy=\chapno\relax\expandafter\stripeq\LABxx%
\else\chapshow{\LAByy}.\expandafter\stripeq\LABxx\fi\fi\fi}%
\ifproofmode\write2{Figure #1}\fi}

\def\tabnum{\global\advance\tableno by 1
\relax\iftablechapternumbers\chapfolio.\fi\the\tableno}

\def\tablabel#1{\global\advance\tableno by 1
\relax\ifproofmode\ifforwardreference
\immediate\write1{\noexpand\expandafter\noexpand\def
\noexpand\csname TABLABEL#1\endcsname{\the\chapno.\the\tableno?}}\fi\fi
\global\expandafter\edef\csname TABLABEL#1\endcsname
{\the\chapno.\the\tableno?}\iftablechapternumbers\chapfolio.\fi
\ifproofmode\llap{\hbox{\marginstyle#1
\kern1.2truein}}\relax\fi\the\tableno}

\def\tabref#1{\hbox{\ifundefined{TABLABEL#1}!!!!
\ifproofmode\ifforwardreference%
\else\write16{ ***Undefined Table Reference #1*** }\fi
\else\write16{ ***Undefined Table Reference #1*** }\fi
\else\edef\LABtt{\getlabel{TABLABEL#1}}%
\def\LABTT{\expandafter\stripchap\LABtt}\iftablechapternumbers%
\chapshow{\LABTT}.\expandafter\stripeq\LABtt%
\else\ifnum\number\LABTT=\chapno\relax\expandafter\stripeq\LABtt%
\else\chapshow{\LABTT}.\expandafter\stripeq\LABtt\fi\fi\fi}%
\ifproofmode\write2{Table#1}\fi}

\def\eq{Eq.~}

\def\fig{Figure~}

\newdimen\sectionskip     \sectionskip=20truept
\newcount\sectno
\def\section#1#2{\sectno=0 \null\vskip\sectionskip
    \centerline{\chaplabel{#1}.~~{\bf#2}}\nobreak\vskip.2truein
    \noindent\ignorespaces}

\def\advancesectno{\global\advance\sectno by 1}
\def\sectfolio{\number\sectno}
\def\subsection#1{\goodbreak\advancesectno\null\vskip10pt
                  \noindent\chapfolio.~\sectfolio.~{\bf #1}
                  \nobreak\vskip.05truein\noindent\ignorespaces}

\def\uttg#1{\null\vskip.1truein
    \ifproofmode \line{\hfill{\bf Draft}:
    UTTG--{#1}--\number\year}\line{\hfill\today}
    \else \line{\hfill UTTG--{#1}--\number\year}
    \line{\hfill\ifcase\month\or January\or February\or March\or April\or
May\or June
    \or July\or August\or September\or October\or November\or December\fi
    \space\number\year}\fi}

\def\contents{\noindent
   {\bf Contents\Z}\nobreak\vskip.05truein\noindent\ignorespaces}

\def\getlabel#1{\csname#1\endcsname}
\def\ifundefined#1{\expandafter\ifx\csname#1\endcsname\relax}
\def\stripchap#1.#2?{#1}
\def\stripeq#1.#2?{#2}

%
\catcode`@=11 
\def\space@ver#1{\let\@sf=\empty\ifmmode#1\else\ifhmode%
\edef\@sf{\spacefactor=\the\spacefactor}\unskip${}#1$\relax\fi\fi}
\newcount\referencecount     \referencecount=0
\newif\ifreferenceopen       \newwrite\referencewrite
\newtoks\rw@toks
\def\refmark#1{\relax[#1]}
\def\refend{\refmark{\number\referencecount}}
\newcount\lastrefsbegincount \lastrefsbegincount=0
\def\refsend{\refmark{\count255=\referencecount%
\advance\count255 by -\lastrefsbegincount%
\ifcase\count255 \number\referencecount%
\or\number\lastrefsbegincount,\number\referencecount%
\else\number\lastrefsbegincount-\number\referencecount\fi}}
\def\refch@ck{\chardef\rw@write=\referencewrite
\ifreferenceopen\else\referenceopentrue
\immediate\openout\referencewrite=referenc.texauxil \fi}
%
{\catcode`\^^M=\active 
  \gdef\obeyendofline{\catcode`\^^M\active \let^^M\ }}%
%
{\catcode`\^^M=\active 
  \gdef\ignoreendofline{\catcode`\^^M=5}}
{\obeyendofline\gdef\rw@start#1{\def\t@st{#1}\ifx\t@st\blankend%
\endgroup\@sf\relax\else\ifx\t@st\bl@nkend\endgroup\@sf\relax%
\else\rw@begin#1
\backtotext
\fi\fi}}
{\obeyendofline\gdef\rw@begin#1
{\def\n@xt{#1}\rw@toks={#1}\relax%
\rw@next}}
\def\blankend{}
{\obeylines\gdef\bl@nkend{
}}
\newif\iffirstrefline  \firstreflinetrue
\def\rwr@teswitch{\ifx\n@xt\blankend\let\n@xt=\rw@begin%
\else\iffirstrefline\global\firstreflinefalse%
\immediate\write\rw@write{\noexpand\obeyendofline\the\rw@toks}%
\let\n@xt=\rw@begin%
\else\ifx\n@xt\rw@@d \def\n@xt{\immediate\write\rw@write{%
\noexpand\ignoreendofline}\endgroup\@sf}%
\else\immediate\write\rw@write{\the\rw@toks}%
\let\n@xt=\rw@begin\fi\fi\fi}
\def\rw@next{\rwr@teswitch\n@xt}
\def\rw@@d{\backtotext} \let\rw@end=\relax
\let\backtotext=\relax

\newdimen\refindent     \refindent=30pt
\def\Textindent#1{\noindent\llap{#1\enspace}\ignorespaces}
\def\refitem#1{\par\hangafter=0 \hangindent=\refindent\Textindent{#1}}
\def\REFNUM#1{\space@ver{}\refch@ck\firstreflinetrue%
\global\advance\referencecount by 1 \xdef#1{\the\referencecount}}
\def\refnum#1{\space@ver{}\refch@ck\firstreflinetrue%
\global\advance\referencecount by 1\xdef#1{\the\referencecount}\refend}

\def\REF#1{\REFNUM#1%
\immediate\write\referencewrite{%
\noexpand\refitem{#1.}}%
\begingroup\obeyendofline\rw@start}
\def\ref{\refnum\?%
\immediate\write\referencewrite{\noexpand\refitem{\?.}}%
\begingroup\obeyendofline\rw@start}
\def\Ref#1{\refnum#1%
\immediate\write\referencewrite{\noexpand\refitem{#1.}}%
\begingroup\obeyendofline\rw@start}
\def\REFS#1{\REFNUM#1\global\lastrefsbegincount=\referencecount%
\immediate\write\referencewrite{\noexpand\refitem{#1.}}%
\begingroup\obeyendofline\rw@start}

\def\REFSCON#1{\REF#1}

\def\cite#1{\refmark#1}
\catcode`@=12 
%
%
%
\proofmodefalse
\baselineskip=13pt plus 1pt minus 1pt
\parskip=2pt
\chapternumberstrue
\forwardreferencefalse
\figurechapternumberstrue
\ifproofmode
\immediate\openout2=allcrossreferfile \fi
\ifforwardreference\input labelfile
\ifproofmode\immediate\openout1=labelfile \fi\fi
\noblackboxes
\hfuzz=1pt
\vfuzz=2pt
 %
 %
\font\tenmib=cmmib10
\font\sevenmib=cmmib10 at 7pt 
\font\fivemib=cmmib10 at 5pt  
\font\tenbsy=cmbsy10
\font\sevenbsy=cmbsy10 at 7pt 
\font\fivebsy=cmbsy10 at 5pt  
\def\BMfont{\textfont0\tenbf \scriptfont0\sevenbf
                              \scriptscriptfont0\fivebf
            \textfont1\tenmib \scriptfont1\sevenmib
                               \scriptscriptfont1\fivemib
            \textfont2\tenbsy \scriptfont2\sevenbsy
                               \scriptscriptfont2\fivebsy}
\def\BM#1{\relax\leavevmode\ifmmode\mathchoice
                      {\hbox{$\BMfont#1$}}
                      {\hbox{$\BMfont#1$}}
                      {\hbox{$\scriptstyle\BMfont#1$}}
                      {\hbox{$\scriptscriptstyle\BMfont#1$}}
                 \else{$\BMfont#1$}\fi}


 %
\def\cropen#1{\crcr\noalign{\vskip #1}}
\newskip\humongous \humongous=0pt plus 1000pt minus 1000pt

\newif\ifdtup

\def\rd{{\rm d}}

\def\gen{{\bf27}}
\def\agen{$\overline{\bf27}$}
\def\sing{{\bf1}}

\def\ten{{\bf10}}

\def\ygen{\gen$^3$}
\def\yagen{\agen$^3$}
\def\ysing{\sing$^3$}
\def\ymix{{\bf 27.$\overline{\bf27}$.1}}

\def\jbar{\bar \jmath}

\def\o{\omega}

\def\slash#1{#1 \kern-.70em /}
\def\del{\partial}
\def\delb{\bar\partial}

\def\y{\eta}

\def\End#1#2#3{{#1}_{\bar\z}{}^{#2}{}_{{#3}\vphantom{\bar\z}}}
\def\splittingtype#1#2{\ifmmode\ca{O}(#1)\oplus\ca{O}(#2)\else
    $\ca{O}(#1)\oplus\ca{O}(#2)$\fi}
\def\real{\Re e}

\def\rd{{\rm d}}
\def\ex#1{{\rm e}^{#1}}
\let\ba=\overline

\let\h=\eta

\def\VEV#1{\left\langle\,{#1}\,\right\rangle}

\def\Y{Yukawa}
\def\T{\ca{T}}
\def\BBW{Bott--Borel--Weil}

\def\ssa{\hbox{\ss a}}

\def\ssA{\hbox{\ss A}}

\def\Cnfg#1#2{\matrix{#1}\mkern-6mu\left[\matrix{#2}\right]}

\def\H#1#2{{\rm H}^{#1}(#2)}

\def\hourandminute{\count255=\time\divide\count255 by
60\xdef\hour{\number\count255}
\multiply\count255 by-60\advance\count255 by\time
\hour:\ifnum\count255<10 0\fi\the\count255}
\def\\{\hfill\break}
\def\:{\kern-1.5truept}

\def\cite#1{{\refmark#1}}

\def\subsection#1{\goodbreak\advancesectno\null\vskip10pt
                  \noindent{\it \chapfolio.\sectfolio.~#1}
                  \nobreak\vskip.05truein\noindent\ignorespaces}

\def\subsubsection#1{\goodbreak\null\vskip10pt
                     \noindent$\underline{\hbox{#1}}$
                     \nobreak\vskip.05truein\noindent\ignorespaces}

\def\bigfract#1#2{{\textstyle#1\over \textstyle#2}}
\def\contents{\line{{\bf Contents}\hfill}\nobreak\vskip.05truein\noindent%
              \ignorespaces}

\long\def\omit#1{}
%
%
\nopagenumbers\pageno=-1
\null\vskip-40pt
\rightline{\eightrm UTTG-01-95, HUB-EP-95/3,
                    HUPAPP-93/5, IASSNS-HEP-93/79}\vskip-3pt
\rightline{\eightrm hepth/9505164}\vskip-3pt
\rightline{\eightrm May 25, 1995}
\vskip .7truein
\centerline{\seventeenrm On the Instanton Contributions to the Masses}
\vskip .2truein
\font\seventeenit=cmmi12 at 17pt
\font\fourteenit=cmmi12 at 14pt
\centerline{\seventeenrm  and Couplings of
 $\hbox{\seventeenit E}_{\hbox{\fourteenit 6}}$ Singlets}
\bigskip\bigskip
\centerline{{\csc P.~Berglund}$^{1}$,\quad
            {\csc P.~Candelas}$^{1,2,4}$,\quad
            {\csc X.~de la Ossa}$^{1}$,}
\vskip2mm
\centerline{{\csc E.~Derrick}$^{2,3}$\footnote{$^{\natural}$}
            {\eightrm Alexander von Humboldt Fellow},\quad
            {\csc J.~Distler}$^{2,4}$\quad{\csc and}\quad
            {\csc T.~H\"ubsch}$^5$\footnote{$^{\flat}$}
            {\eightrm On leave from the Institute ``Rudjer
              Bo\v{s}kovi\'c'', Zagreb, Croatia.}}
\bigskip \bigskip
\centerline{
\vtop{\baselineskip=12pt\hsize = 2.0truein
\centerline{$^1$\it School\:\ of\:\ Natural\:\ Sciences}
\centerline{\it Institute\:\ for\:\ Advanced\:\ Study}
\centerline{\it Olden Lane}
\centerline{\it Princeton, NJ\:\ 08540}}\quad
\vtop{\baselineskip=12pt\hsize = 2.0truein
\centerline{$^2$\it Theory Group}
\centerline{\it Department of Physics}
\centerline{\it University of Texas}
\centerline{\it Austin, TX 78712}}\quad
\vtop{\baselineskip=12pt\hsize = 2.0truein
\centerline{$^3$\it Humboldt Universit\"at zu Berlin}
\centerline{\it Institut f\"ur Physik}
\centerline{\it Invalidenstrasse 110}
\centerline{\it D-10115 Berlin, Germany}} }
\bigskip
\centerline{
\vtop{\baselineskip=12pt\hsize = 2.0truein
\centerline{$^4$\it Joseph\:\ Henry\:\ Laboratories}
\centerline{\it Jadwin Hall}
\centerline{\it Princeton University}
\centerline{\it Princeton, NJ\:\ 08544}}\quad
\vtop{\baselineskip=12pt\hsize = 2.0truein
\centerline{$^5$\it Department of Physics}
\centerline{\it Howard University}
\centerline{\it 2355 6th St. NW}
\centerline{\it Washington, DC 20059} }}
\bigskip \bigskip
\vbox{\centerline{\bf ABSTRACT}
\vskip.2truein
\vbox{\baselineskip 12pt\noindent  We consider
the gauge neutral matter in the low--energy
effective action for string theory compactification
on a \cym\ with $(2,2)$ world--sheet supersymmetry.
At the classical level these states (the \sing's of
$E_6$) correspond to the cohomology group $H^1(\M,{\rm End}\>T)$. We
examine the first order contribution of instantons to the mass matrix of
these
particles.  In principle, these corrections depend on the \K\ parameters
$t_i$
through factors of the form $e^{2\p i t_i}$ and also depend
on the complex structure parameters. For simplicity we consider in
greatest
detail the quintic threefold $\cp4[5]$.
It follows on general grounds that the total mass is often, and
perhaps always, zero. The contribution of individual instantons is
however nonzero and the contribution of a given instanton may develop
poles associated with instantons coalescing for certain values of the
complex structure. This can happen when the underlying \cym\ is
smooth. Hence these poles must cancel between the coalescing
instantons in order that the superpotential be finite.
We examine also the \Y\ couplings involving neutral
matter \ysing\ and
neutral and charged fields \ymix, which have been little investigated even
though
they are of phenomenological interest.  We study the general
conditions under which these couplings vanish classically.  We also
calculate
the first--order world--sheet instanton correction to these couplings
and argue that these also vanish.}}
\newpage
\contents
\bigskip
\item{1.~}Introduction
\medskip
\item{2.~}Review of What is Known Classically
\itemitem{\it 2.1~}{\it Deformations of the complex structure and of the
tangent bundle}
\itemitem{\it 2.2~}{\it The number of singlets}
\itemitem{\it 2.3~}{\it Integrability and the vanishing of the \ysing\
coupling}
\itemitem{\it 2.4~}{\it The vanishing of the \ymix\ coupling}
\medskip
\item{3.~}Preliminaries Concerning Instanton Contributions to Correlation
Functions
\itemitem{\it 3.1~}{\it The zero modes}
\itemitem{\it 3.2~}{\it Vertex operators and powers of $g$}
\itemitem{\it 3.3~}{\it Decomposition of singlets along a line}
\medskip
\item{4.~}Masses Generated by Instantons
\itemitem{\it 4.1~}{\it The mass matrix}
\itemitem{\it 4.2~}{\it Other splitting types}
\medskip
\item{5.~}{Examples of the Effects of Lines in $\cp4[5]$}
\itemitem{\it 5.1~}{\it General facts}
\itemitem{\it 5.2~}{\it An \splittingtype{0}{-2} line as the limit of
                        \splittingtype{-1}{-1} lines}
\itemitem{\it 5.3~}{\it The singlet mass matrix is probably always zero}
\medskip
\item{6.~}The Yukawa Couplings
\itemitem{\it 6.1~}{\it The couplings in terms of zero modes}
\itemitem{\it 6.2~}{\it The form of the Yukawa couplings}
\medskip
\item{7.~}Conclusion
\itemitem{\it 7.1~}{\it Summary and discussion}
\itemitem{\it 7.2~}{\it Open questions}
\medskip
\item{A.~}{Remarks on the Determinants and  Resolution of
an Apparent Paradox}
\newpage
\headline={\ifproofmode\hfil\eightrm draft:\ \today\
\hourandminute\else\hfil\fi}
\pageno=1\footline={\rm\hfil\folio\hfil}
\section{intro}{Introduction}
\smallskip
\noindent \cy\ compactifications correspond to
string theory vacua with $(2,2)$ world--sheet
supersymmetry.
In these theories, there is now a good understanding
of many aspects of the low--energy effective theories
that correspond to these vacua.
The families (\gen's) and anti-families (\agen's)
of $E_6$ are in one--one correspondence with the complex
structure and \K\ class parameters of the manifold.
The geometry of the parameter spaces is coming
to be understood, and, in virtue of mirror symmetry,
the \ygen\ and \yagen\ \Y\ couplings may be calculated
exactly in the sigma model.
These \Y\ couplings may be expressed,
equivalently, in terms of the periods
of the holomorphic three--form over the manifold~
\REFS{\AS}{A.~Strominger, \cmp{133} (1990) 163.}
\REFSCON{\Cd}{P.~Candelas and X.~C.~de~la~Ossa, \npb{355} (1991) 455.}
\refsend\
or in terms of a convergent instanton expansion.
There are, however, other parameters in these models \
\Ref{\witten}{E.~Witten, \npb{268} (1986) 79.}\
which correspond to $E_6$ singlets, \sing.
Geometrically, these parameters correspond to
deformations of the tangent bundle of the \cym.
The infinitesimal deformations of the tangent bundle
correspond to the cohomology group $H^1(\M,{\rm End}\>\T)$,
which is generically nontrivial.
Unfortunately, these parameters are of a more recondite character
than is the case for the \gen's and \agen's
and, until recently~~\Ref\SW{E.~Silverstein and E.~Witten: ``Criteria
for Conformal Invariance of $(0,2)$ Models'', hepth/9503212.},
little was known about whether they acquire mass
by non--perturbative corrections or about the \Y\ couplings,
the \ymix\  and \ysing , into which they enter.
These couplings are of
phenomenological interest \
\REF{\ross}{G.~G.~Ross, \plb{221} (1988) 315.}
\cite{{\witten,\ross}},
and our lack of understanding
constitutes a considerable barrier to model building.
An improved understanding of these parameters
would also aid in the exploration of the largely
mysterious class of $(0,2)$ vacua, which are potentially
of considerable phenomenological importance.
Quite apart from any phenomenological considerations, the
\ygen\ and \yagen\ \Y\ couplings are of interest because they reflect
deep geometrical properties of the \cym\ and of
its moduli space and it seems likely that the same should be true of the
couplings that involve the singlets.

In the absence of any insight into the geometry of the parameter space
corresponding to the singlets we focus, in this article, on the much more
modest
goal of studying the instanton contribution to the mass matrix of the
singlets
and the \ymix\  and \ysing\  \Y\ couplings following the
techniques of early papers on the instanton contributions to
the \ygen\ and \yagen\ couplings~
\REF{\DSWW}{M. Dine, N. Seiberg, X.-G. Wen and E. Witten,\\
\npb{278} (1986) 769, {\bf B289} (1987) 319.}
\REF{\DG}{J.~Distler and B.~Greene, \npb{309} (1988) 295.}
\cite{{\DSWW,\DG}}.
To do the calculations, we follow\
\cite{\DG}\
by writing down the path integral
corresponding to the Yukawa coupling, which involves
an insertion of three appropriate vertex operators:
 $$
y_{ijk} = \frac{\int {\cal D}[x] \, V_i V_j V_k e^{-S} }
{ \int {\cal D}[x] \,e^{-S}}  \quad .
\eqlabel{genyukawa}$$
The integral breaks up into a sum over saddle points of
different instanton degree; these correspond to
the topologically different ways the world--sheet can be mapped into
the target space.
This leads to an expansion:
 $$\eqalign{
\int {\cal D}[x] \, V_i V_j V_k e^{-S} =&
   e^{-S_{free}\vert_0} \int {\cal D}[\tilde{x}] \,
      V_i V_j V_k e^{-S_{int}} \Big\vert_0 \cr
& + ~  e^{-S_{free}\vert_1} \int {\cal D}[\tilde{x}] \,
      V_i V_j V_k e^{-S_{int}} \Big\vert_1 \quad + \quad  \ldots \cr }
\eqlabel{genexpansion}$$
The integrals that remain are over the modular group of the world--sheet
and the zero modes of the instanton.

The mass matrix arises from an
$\bf R.1.1$ coupling, where $\bf R$ denotes the ``dilaton'' (the scale of
the
internal manifold) and so is,
in reality, a Yukawa coupling also.  There is also a $\bf C.1.1$
coupling,
where $\bf C$ is the vertex operator for a complex structure modulus,
that is related by mirror symmetry to $\bf R.1.1$ of the mirror manifold.
Logically\Footnote{We are grateful to E.~Witten for stressing this point
to
us.} the study of the mass matrix precedes the study of the couplings
\ymix\  and \ysing, since those singlets that acquire mass are absent
from the low energy theory and the couplings \ymix\  and \ysing\ into
which they enter are irrelevant. Furthermore we can only calculate
on-shell
quantities in string theory. The superpotential couplings, that is
zero-momentum
2-- and 3--point couplings of the chiral fields, are on-shell {\it
provided}
the
fields are massless. If we find that a certain singlet is massive (appears
quadratically in the superpotential), then the string calculations of its
3--
and
higher--point couplings are off-shell, and hence ambiguous.

This investigation began with a computation of the instanton
corrections to the \ysing\ and \ymix\ couplings. The resulting
expressions fail to vanish owing to certain $\d$-function
contributions that arise on integrating by parts. The resulting
couplings are ambiguous in that they are ill-defined in BRST
cohomology. It appeared that there were two possible resolutions to
this. On the one hand the singlet could acquire mass through instanton
corrections in which case one would expect the couplings to be
ill-defined in BRST cohomology and they would also be irrelevant since
they would not be couplings between massless particles. A second
possibility is that the ambiguous contributions should be removed by a
proper accounting for contact terms in the conformal field theory.

In studying the the mass matrix for the \sing's we find that the
individual instantons contribute amounts that are generically
non-zero. A general argument following~~\cite{\SW},
however implies that these must cancel in the sum. Moreover all
contributions to the superpotential that involve the \sing's should
vanish. The consequence for the  couplings is that the ambiguous
contributions to the \Y\  couplings should be canceled by contact
terms. In fact once we know that the masses vanish, the particular
form of our expressions is such that the only way to ascribe
unambiguous meaning to them is that they should indeed vanish.

We find the individual contributions of the instantons to be rather
complicated.
For example each instanton, $L$, makes a contribution to the mass matrix
that
depends
on the complex structure and \K\ parameters of the manifold {\it and\/}
on the particular instanton~$L$. For the \yagen\ coupling, by contrast,
the coupling depends only on the \K\ parameters of the manifold and, in
each
degree, each instanton contributes equally. That is, it is sufficient
to know the number of instantons of each degree, and it is
not necessary to know
the location of each instanton. For the singlet couplings, however,
much more detailed information is
necessary.
As the complex structure parameters of the manifold are varied,
the instantons move in the manifold; though if
properly counted, their number remains fixed.
Thus a zero total mass is achieved by very complicated cancellations
between the contributions of the different instantons. This situation
is of potential interest to mathematicians; it is possible to ascribe
a matrix, computed from $H^1(\M,{\rm End}\>\T)$, to each instanton in
such a way that the sum over all instantons vanish. It is clear that
this persists to all orders  in the instanton expansion.

The layout of this paper is the following. In Section~2 we review what is
known about the classical values of the \ysing\ couplings and
we show that the \ymix\ couplings vanish whenever the singlet corresponds
to a polynomial deformation of the tangent bundle.
In Section~3 we gather together the basic
elements that we need to calculate the instanton corrections to the
couplings.
These are the zero modes of the fields about the instanton and the form of
the
vertex operators that we will need. A basic technique is the decomposition
of
tensors into components that are sections of line bundles over the
instanton
and the application of the Bott--Borel--Weil theorem which gives the
cohomology of forms that take values in these bundles. We then discuss, in
Section~4, the form of the mass matrix for the singlets. In particular
we consider the
mass matrix due to the simplest type of instanton, which has normal bundle
\hbox{$\ca{O}(-1)\oplus\ca{O}(-1)$}. This is in some sense the generic
case
but other cases are possible and indeed arise as the parameters of the
\cym\
are varied. It can happen, for example,
that for special values of the parameters,
two (or more) instantons coincide and the normal bundle of each
degenerates to
\hbox{\splittingtype{0}{-2}}.
In this situation the contribution of each line to the mass matrix has
a pole; though the poles cancel between the coalescing lines.
We examine this situation in Section~5
in the context of the manifold $\cp4[5]$ and extend the argument
of~~\cite{\SW} to show that the total mass and all the couplings
involving \sing\ vanish for a wide variety of situations.
We turn, in Section~6, to a detailed computation  of the first order
instanton corrections to the Yukawa
couplings \ysing\ and \ymix\ and a discussion of the resulting expression.
In Section~7 we discuss our results and list a number of
open questions.
Finally, an appendix deals with the determinants associated to
expansion about the instanton and issues related to the zero modes
about multiple instantons that result from the coalescence of isolated
instantons.
\newpage
\section{classical}{Review of What is Known Classically}
\vskip-20pt

\subsection{Deformations of the complex structure and of the tangent
bundle}
At least some (and in favourable circumstances all) deformations of the
tangent bundle $\T$ on the manifold $\cal M$ may be described using the
Kodaira-Spencer deformation theory~~\Ref{\rKodaira}{K.~Kodaira,
{\sl Complex Manifolds and Deformations of Complex Structure}\\
(Springer Verlag 1985).} {}~as presented, for example, in Refs.~
\REFS{\rPhilip}{P.~Candelas, \npb{298} (1988)
458.}\REFSCON{\rGSW}{M.B.~Green,
J.H.~Schwarz and E.~Witten,
 {\it Superstring Theory, Vol.~I and II}\hfill\break (Cambridge University
Press, Cambridge, 1987).}
\refsend.
A vector tangent to a projective space may be understood as a differential
operator $v^A{\partial\over\partial z^A}$ that acts on functions
homogeneous
of degree zero. Since
 $$
z^A \pd{f}{z^A} = \ell f
 $$
for a function homogeneous of degree $\ell$ we may
make the identifications
 $$
v^A\sim v^A+z^A~. \eqlabel{vident}
 $$
In order to be tangent to the hypersurface $p^\a=0$ the vector must
satisfy
 $$
v^A{\partial p^\a\over\partial z^A}=0~, \eqlabel{Tdef}
 $$
which is compatible with \eqref{vident} since
$z^A{\partial p^\a\over\partial z^A}={\rm deg}(\a)p^\a=0$ on $\ca{M}$. A
deformation of this structure may be realised by replacing equation
\eqref{Tdef} by
 $$
v^A\left({\partial p^\a \over\partial z^A}+\ssa^\a_A\right)=0
\eqlabel{defm}
 $$
where the $\ssa^\a_A$ form a set of
polynomials that are subject to the constraint
 $$ z^A \ssa^\a_A(z)=0~. $$
It is then natural to set
 $$
a^\m{}_\n=-{1\over 2\p i}\,\ssa^\a_\n \ch^\m{}_{\bar\r\a}dx^{\bar\r}~,
\eqlabel{polend}
 $$
where $\ssa^\a_\n=\ssa^\a_A{\partial z^A\over\partial x^\n}$ is the
projection of $\ssa^\a_A$ along $\ca{M}$. This
gives an explicit representation of the elements of
$H^1_{\bar\partial}(\ca{M},{\rm End}\>\T)$ in terms of the polynomials
$\ssa^\a_A$; a factor of $-1/2\p i$ will simplify later expressions.
The quantity $\ch^\m{}_{\bar\r\a}$ is
the extrinsic curvature
 $$\eqalign{
\ch_{\m\n}{}^\a &=
\pd{z^a}{x^\m} \pd{z^b}{x^\n} \left( \ppd{p^\a}{z^a}{z^b} -
\G_{ab}^c \pd{p^\a}{z^c} \right) \cropen{5pt}
&= \pd{z^A}{x^\m} \pd{z^B}{x^\n} \ppd{p^\a}{z^A}{z^B}\cr}
 $$
where the $z^a$ are coordinates for the embedding space corresponding
to taking $z^5=1$, say. In passing to the second equality we use the
fact that in the Fubini-Study metric the term containing the
connection does not contribute.
The extrinsic curvature is perhaps more familiar from its occurrence
in the representation
 $$
h^\m=-{1\over 2\p i}\,q^\a\ch_{\bar\n}{}^\m{}_\a dx^{\bar\n}~,
 $$
where elements of  $H^1_{\bar\partial}(\ca{M},\T)$ are represented in
terms
of the polynomials $q^\a$~\cite{{\rPhilip,\rGSW}}.

\subsection{The number of singlets}
In the previous section we found a very convenient way of representing
(some of) the~\sing's. However, as for the moduli fields associated to the
{\bf27}'s, one needs to consider the full Koszul complex and its
corresponding
spectral sequence in order to obtain the full group $H^1(\M,{\rm
End}\>T)$.
Though, unlike for the {\bf27}'s and the $\overline{\bf27}$'s, the number
of {\bf1}'s
is {\it not} constant over the space of complex structures~(see
Ref.~\REFS{\refsings}{P.~Berglund, T.~H\"ubsch and L.~Parkes,
Mod. Phys. Lett. {\bf148} (1990) 1485;}
\REFSCON{\rMatter}{P.~Berglund, T.~H\"ubsch and L.~Parkes,
           \cmp{148} (1992) 57.}
\refsend)---even at the
classical level. For certain special choices of the complex structure
parameters, the number of the $E_6$ {\bf1}'s which correspond to elements
of
${\rm H}^1(\M,{\rm End}\>\T)$ is larger than for a generic choice of these
parameters.

As an example of this phenomenon let us consider the following \cy\
manifold~\REFS{\rrolf}{R.~Schimmrigk: \plb{193} (1987)175.},
$$
  {\cal M} \in \Cnfg{\cp3\cr \cp2\cr}{3&1\cr 0&3\cr}~~ :
 ~~\left\{
\eqalign{
   f(x)   &= ~\sum_{i=0}^3 x_i^3\cr
   g(x,y) &= ~\sum_{i=1}^3 x_i y_i^3 + a x_0 y_1y_2y_3\cr}
\eqalign{
&\vphantom{\sum_{i=0}^3}= ~0~.  \cr
&\vphantom{\sum_{i=0}^3}= ~0~,  \cr}\right.
\eqlabel{cicyR}
$$
For a generic choice of $f,g$, and in particular for the above defining
equations with $a\neq 0$,
one can show that $\dim H^1(\M,{\rm End}\>\T)=88$. However, when $a=0$
the number of elements of ${\rm H}^1(\M,{\rm End}\>\T)$ jumps to $108$. In
the
explicit computation this is because certain maps vanish in the
long exact cohomology sequence  in which ${\rm H}^1(\M,{\rm End}\>\T)$
is an element. (For more details, see pp.240
in~
\Ref{\rBeast}{T.~H\"ubsch, {\it \cy\ Manifolds---A Bestiary for
Physicists}\Z (World Scientific, Singapore, 1992).}.)

Evidently, the number of massless {\bf1}'s can jump as we vary the the
complex structure. It also turns out that the number can jump as we vary
the K\"ahler structure, though the extra singlets which arise in that case
have nothing to do with the classical $H^1(\M,{\rm End}\>\T)$.
In many cases, one can understand this jump in the number of massless
{\bf1}'s as being due to symmetries. For instance, Landau-Ginzburg models
possess a discrete R-symmetry in spacetime (corresponding to the quantum
symmetry of the Landau-Ginzburg orbifold). In many such theories, the
number of
massless {\bf1}'s jumps at the Landau-Ginzburg locus. The existence of the
extra massless {\bf1}'s (which arise in twisted sectors, and hence
transform non-trivially under the ``quantum" discrete R-symmetry) can be
understood very simply~
\Ref{\rJDSK}{J.~Distler and S.~Kachru, \plb{336} (1994),
                                       {\tt hep-th/9406091}.}.\
Were they not present, the R-symmetry would suffer from a gravitational
anomaly in spacetime.

It should be emphasized that the number of massless {\bf1}'s does not
always jump at the
Landau-Ginzburg locus -- $\cp4[5]$ is an example where the number does
not jump -- but in those cases, the R-symmetry is non-anomalous without the
need for extra {\bf1}'s. Whenever the R-symmetry would be anomalous
without them, however, extra massless {\bf1}'s appear at the
Landau-Ginzburg locus to cancel the anomaly.

In the handful of explicitly worked examples, including the one
discussed above, the jump in the number of massless {\bf1}'s as one varies
the complex structure also seems to be associated with occurrence of
discrete $R$-symmetries. In \eqref{cicyR},
for $a=0$, there is a symmetry $R:\,x_0\to\alpha x_0\,,\alpha^3=1$. (This
is
an $R$-symmetry since the holomorphic $3$-form transforms non-trivially.)
It seems likely that cancelling the would-be-anomaly in the discrete
R-symmetry is the  ``explanation" for why the extra massless {\bf1}'s
appear in all these cases
\Footnote{Note, again, that a discrete R-symmetry {\it by itself} does not
imply the existence of extra massless {\bf1}'s; the symmetry could be
non-anomalous without additional massless fields. For example, in the
$\cp4[5]$ family of models, the number of {\bf1}'s is constant,
regardless of $R$-symmetries.}. It is not clear, however, that this one
mechanism will account for all cases where the number of massless {\bf1}'s
jumps. Perhaps there are other, as yet undiscovered, mechanisms at work.

Finally, it should be clear that a physically complete moduli space for
\cy\
compactification\Footnote{In general, it is only natural to consider the
(2,2)-supersymmetric ``standard'' \cy\ Ansatz merely as a special subset
of
the more general (0,2)-supersymmetric framework.} must be spanned by the
moduli corresponding to the {\bf27}'s and $\overline{\bf27}$'s, and all
the
massless {\bf1}'s with exactly flat potential.

What are we to make of the extra massless {\bf1}'s which arise on certain
codimension-1 loci in the (2,2) moduli space? In most cases, one expects
that they, though massless, do not  (even classically) have a flat
superpotential. In the language of the next subsection, they represent
infinitesimal, but not integrable, deformations. In some cases, however,
it is possible that the locus in question is a multicritical point, where
two different branches (of, possibly, different dimensions) of the moduli
space meet. In any case, we will restrict ourselves to one branch of the
moduli space, and so focus on those \sing's which are (classically)
massless for all
values of the complex structure.

\subsection{Integrability and the vanishing of the ${\bf 1}^3$ coupling}
Elements of $s^{\mu}{}_\nu\in H^1(\M,{\rm End}\>T)$
correspond to first order deformations of the holomorphic structure of the
tangent bundle to $\M$ .
Such a deformation can be thought of  \cite{\rKodaira} as defining a
deformed $\bar\del$-operator,
$$
(\bar D)^\mu{}_\nu=\delta^\mu{}_\nu \bar\del
+s^{\mu}{}_\nu(\epsilon)\wedge
\eqlabel{Ddef}
$$
acting on sections of $\> T$. $s^{\mu}{}_\nu(\epsilon)$ can be expanded in
powers of $\epsilon$,
$$s(\epsilon)=\epsilon^i s_i + \epsilon^i \epsilon^j s_{ij} +\dots
\eqlabel{sexp}
$$
Demanding that $\bar D^2=0$, one finds, to first order in $\epsilon$, that
$(s_i)^\mu{}_\nu\in H^1(\M,{\rm End}\>T)$. To second order, one finds the
condition,
$$\bar\del s_{ij} + [s_i,s_j]=0$$
(where $[\cdot,\cdot]$ means commutator as elements of ${\rm End}\>T$, and
wedge-product as (0,1)-forms, and hence is symmetric in its arguments.)
In order for the deformation to exist to second order, $[s_i,s_j]$ must be
$\bar\del$-exact. At each order in $\epsilon$ one finds a new potential
obstruction, which must be trivial as an  element of $H^2(\M,{\rm
End}\>T)$ for the deformation to be integrable to that order.

If {\it all} of the obstructions vanish, then we say that the deformation
is integrable, and, in that case,  it is possible to define a covariant
derivative on
the space of parameters of the complex structure of the tangent bundle.
It  can then be shown
{}~~\Ref\CD{P.~Candelas and X. de la Ossa, unpublished.}~
that the singlets can be represented as derivatives of a background
gauge field,
  $$
s_i={\cal D}_i a~
  $$
and that
$$
\bar\del ({\cal D}_i s_j)=[s_i,s_j]~.
\eqlabel{sseq}
 $$

Contracting the fermions, one readily sees that the classical contribution
to the Yukawa coupling is
$$
y(s_i,s_j,s_k)=
\int_{\M}\Omega\wedge {\rm Tr} (s_i[s_j,s_k])
\eqlabel{sss}
 $$
 From the above analysis, one sees that this vanishes precisely when the
deformation is integrable to second order~\cite{{\witten,\rGSW}}.
Presumably this correspondence persists to higher order, if the
deformation is integrable to $n^{th}$ oder, then the classical
contribution to the spacetime superpotential vanishes through order
$(n+1)$, and vice versa.

Singlets corresponding to polynomial deformations are, manifestly,
integrable. Hence, it comes as no surprise that,
using the tensor formalism developed
in~\cite{{\rMatter,\rBeast}},
one can show directly that if the three $\bf 1$'s correspond to polynomial
deformations then the \ysing\ coupling vanishes. In particular, for
\cp4[5]
all $E_6$ singlets are polynomial deformations and, as discussed
previously, they may be thought of as the deformations of the 1st
differential
of the defining polynomial, which are not 1st differentials of the
deformations
of the defining polynomials~\cite{\rBeast}:
$$\vth(x) \sim \d \rd f(x) = \d \rd x^a f_{abcde}x^bx^cx^dx^e
                   = \rd x^a \vth_{a(bcde)}x^bx^cx^dx^e~. $$
They are therefore represented by the tensor $\vth_{a(bcde)}$, which is
symmetric in $bcde$, but vanishes upon symmetrization on all five indices.
Next note that the ${\bf1}^3$ coupling requires one holomorphic 3-form,
represented by one $\e$-tensor. The cubic coupling is therefore the
product
$$ \e_{a_1b_1c_1d_1e_1}\>\vth^{(1)}_{a_2b_2c_2d_2e_2}\,
        \vth^{(2)}_{a_3b_3c_3d_3e_3}\,\vth^{(3)}_{a_4b_4c_4d_4e_4}~, $$
which is impossible to make into an invariant. This may be seen in a
number of
ways the simplest being that the five indices of
$\e_{a_1b_1c_1d_1e_1}$ must be contracted with indices of different
$f^{ab\ldots c}$'s, at least five $f^{ab\ldots c}$'s must be used. This
however
provides 25 superscripts, to be contracted with 20 subscripts which is
impossible, and so the coupling has to vanish.

Distler and Kachru~~\Ref\DK{J. Distler and S. Kachru,
\npb{430} (1994) 13, hep-th/9406090.}\ have shown that, for
arbitrary choice of defining polynomial $W(x)$, the singlets correspond to
200
exactly flat directions in the spacetime superpotential, at the
Landau-Ginzburg point.  These flat directions break (2,2) worldsheet
supersymmetry to (0,2), but it is not immediately clear whether these flat
directions persist {\it away} from the Landau-Ginzburg point.
Distler and Kachru gave an indirect argument that this {\it is} the case,
and this claim is further explored by Silverstein and
Witten~~\REF{\Silver}{E.~Silverstein, ``Miracle at the Gepner
Point'', hep-th/9503150.}\cite{{\SW,\Silver}} who argue that in fact the
remaining 24 \sing's
are associated with flat directions as well.
The present paper can be seen as an attempt to test the hypothesis from
the
opposite, \cy\ (large radius) phase, by probing the instanton corrections
to the singlet superpotential.

\subsection{The vanishing of the $\bf 27.\overline{27}.1$
coupling}
We have seen above that whenever the classical part of the \ysing\
coupling
vanishes, the deformations of $H^1_{\bar\partial}(\ca{M},{\rm End}\T)$ are
integrable. It is tempting to speculate, based on this, that the classical
part of the \ymix\ coupling should also vanish. However in this case we do
not
have a deformation-theoretic interpretation of the coupling analogous to
the
previous case so we limit ourselves to a discussion of simple cases for
which
the \ymix\ coupling can be shown to vanish, and the limitations of these
arguments.

One of the elements of $H^1_{\bar\partial}(\ca{M},\T^\ast)$ that is always
present is the one corresponding to the K\"ahler form on $\cal M$ itself,
 $$
b_\n=g_{\n\bar\s}dx^{\bar\s}~.
 $$
If in addition the $\bf 1$ also corresponds to a polynomial deformation
then
we find
 $$\int_{\ca{M}} \O\wedge h^\m\wedge b_\n\wedge s^\n{}_\m
=-\int_{\ca{M}}\O\wedge h^\m
\ssa^\a_\m\,\ch_{\bar\r\bar\s}dx^{\bar\r}\wedge dx^{\bar\s}=0~,
 $$
the last equality being due to the fact that $\ch_{\r\s}{}^\a$ is
symmetric
in its lower indices. Thus the coupling vanishes for the case that the
$\bf 1$
corresponds to a polynomial deformation and the \agen\  corresponds to the
\K-form on $\cal M$. However, the \K-form on $\cal M$ may be written as
$b_\n=\sum_iv^i\o_{\n,i}$, where $\o_{\n,i}$ form a basis for
$H^1_{\bar\partial}(\ca{M},\T^\ast)$ and $v^i$ are real parameters subject
to
some (finite number of) open conditions such as to form the K\"ahler cone.
Therefore (with some abuse of notation),
 $\sum_i v^i\,({\bf27}.\o_{\n,i}.{\bf1})=0$. On the other hand, the $v^i$
are
linearly independent, hence $({\bf27}.\o_{\n,i}.{\bf1})=0$ for all $i$.

Thus, all classical $\bf 27.\overline{27}.1$ Yukawa couplings vanish
whenever
the \sing's can be represented by polynomial deformations. Moreover, for
the
special choice of the \agen\ which corresponds to the large radius limit
(opposite to the Landau-Ginzburg point), these mixed Yukawa couplings
vanish
exactly.

\newpage
\section{general}{Preliminaries Concerning Instanton Contributions to
Correlation Functions}
\vskip-20pt
\subsection{The zero modes}
The sigma-model action at string tree level is given by
 $$\eqalign{
S~=~\int d^2z\Big\{
g_{\m\bar\n}(\del_z X^\m \del_{\bar z} X^{\bar\n}&+\del_{\bar z} X^\m
\del_z
X^{\bar\n}) +
\ps_{\bar\m}\big[ \del_{z}\ps^{\bar\m} + \G^{\bar\m}_{\bar\n\bar\r}
(\del_{z}X^{\bar\n})\ps^{\bar\r} \big]\cr
&+ \l_\n\big[\del_{\bar z}\l^\n + \G^\n_{\r\s}(\del_{\bar z} X^\r) \l^\s
\big]
+ R^\m{}_\n{}^{\bar\r}{}_{\bar\s}\l_\m\l^\n\psi_{\bar\r}\psi^{\bar\s}
\Big\}~.
\cr}
\eqlabel{saction}$$

The instantons correspond to mappings such that
 $$
\del_{\bar z} X^\m~=~0~~~~,~~~~\del_{z} X^{\bar\n}~=~0
\eqlabel{binstanton} $$
and
 $$
\eqalign{
\del_{\bar z}{\bar\ps}^{\m}~&=~0~,\cr
\del_{\bar z}{\bar\ps}_{\n}~&=~0~,\cr}
\hskip70pt
\eqalign{
\del_{\bar z} \l^\m~&=~0~,\cr
\del_{\bar z} \l_\n~&=~0~.\cr}  \eqlabel{finstantons} $$
where we have conjugated the $\ps$ equations to make the point that we are
seeking
elements of the Dolbeault groups $H^0(L,\ca{S}\otimes\ca{T}_L)$ and
$H^0(L,\ca{S}\otimes\ca{T}^\ast_L)$, in which $\ca{S}$ denotes the spin
bundle and $\ca{T}_L$ and $\ca{T}^\ast_L$ are the holomorphic tangent and
cotangent bundles of $L$.
Note that the fermionic equations do not contain terms involving the
connection in virtue of the bosonic equations \eqref{binstanton}. The
bosonic
equations are the statement that the embedding of the worldsheet in the
manifold is holomorphic. In the neighbourhood of the instanton we may
choose local coordinates on the manifold such that $X^3=\z$ is along the
direction of the instanton, and $\x$ and $\y$ are sections of the normal
bundle, as indicated in \fig\figref{finst}.
\midinsert
\iffigmode
\def\figinst{\hbox{\epsfxsize=4.5truein
                    \epsfbox{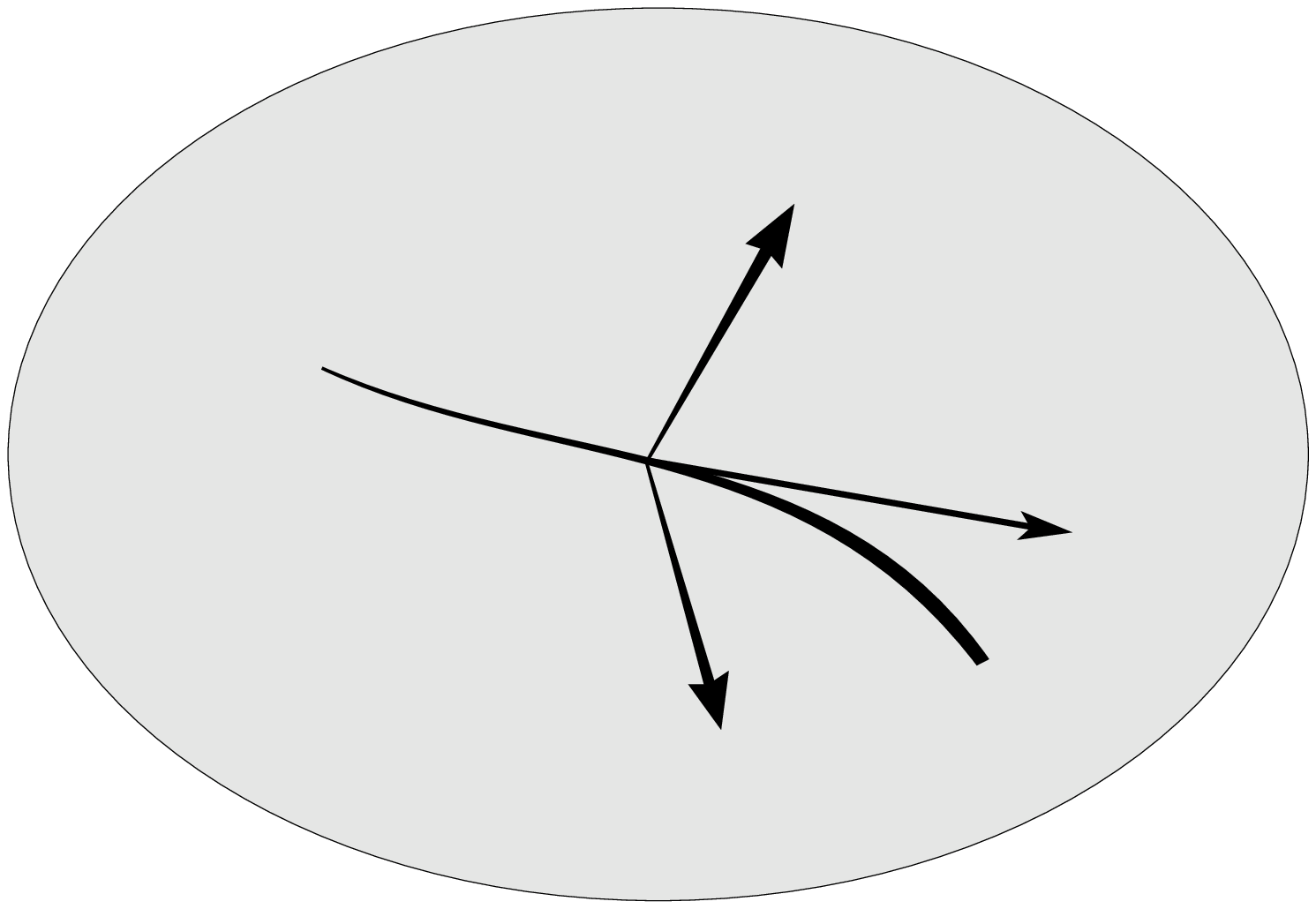}}}
\figbox{\figinst\vskip15pt}{\figlabel{finst}}
{The coordinates on the Calabi--Yau manifold \M\ in the
neighbourhood of the instanton $L$.}
\place{1.4}{2.25}{$\fourteenmath\M$}
\place{2.5}{3.0}{$\fourteenmath L$}
\place{4.8}{2.5}{$\fourteenmath \z$}
\place{3.8}{3.8}{$\fourteenmath \x$}
\place{3.5}{1.5}{$\fourteenmath \y$}
\else
\figbox{{\vrule height 3.5truein width 0pt}\vskip15pt}{\figlabel{finst}}
{The coordinates on the Calabi--Yau manifold \M\ in the
neighbourhood of the instanton $L$.}
\fi\bigskip
\endinsert
Now it is standard analysis that the tangent bundle of the manifold
decomposes
into the tangent bundle and the normal bundle to the instanton
 $$
\ca{T}~=~\ca{T}_L\oplus\ca{N}$$
and owing to the fact that the instanton is
a holomorphic submanifold this decomposition is holomorphic. Furthermore
any
holomorphic vector bundle over a sphere decomposes holomorphically into a
direct sum
of line bundles,
$\ca{O}(k)$, that are classified by integers, $k$, corresponding to their
first
Chern class. For the tangent bundle to $L$ we have
 $$
\ca{T}_L~=~\ca{O}(2)$$
since $c_1(\ca{T}_L)=\ch(S^2)=2$. Decomposing also the normal bundle
$\ca{N}_L$ we have
 $$
\ca{T}~=~\ca{O}(p)\oplus\ca{O}(q)\oplus\ca{O}(2)~.$$
Since also $c_1(\ca{T})=0$, in virtue of the fact that the manifold is
\cy, we
have also that
 $$
p+q~=~-2~.$$
Various values of $p$ and $q$ are possible though for most manifolds
$p=q=-1$
is the generic case, in the sense that for a generic choice of parameters
all
the lines are discrete (they do not lie in continuous families) and have
normal bundle \splittingtype{-1}{-1}. The issues of discreteness and
splitting
type are intimately related since a line which has splitting type other
than
\splittingtype{-1}{-1}\ has deformations, \ie it moves in a continuous
family
at least infinitesimally. We shall first take the normal bundle to be
\splittingtype{-1}{-1} and for this case the map from the worldsheet to
the
instanton is of the form
 $$
X^3=\z={az + b\over cz + d}\qquad,\qquad
\pmatrix{a&b\cr c&d\cr} \in SL(2,\IC) \qquad.   \eqlabel{inst.eq}
 $$

Returning to equations \eqref{finstantons} we now need to know only that
$\ca{S} = \ca{O}(-1)$ and that the dimension of $H^0(S^2,\ca{O}(k))$ is
given
as a special case of the \BBW\ theorem by
 $$
\hbox{dim}H^0(S^2,\ca{O}(k))~=~
\cases{k+1~,\qquad &for $k\geq  0 $ \cropen{5pt}
         0~,       &for $k\leq -1~$. \cr} $$
For the $\l^\m$ equation, for example, we are concerned with
 $$
\ca{S}\otimes\ca{T}
{}~=~\ca{O}(-1)\otimes\Big(\ca{O}(-1)\oplus\ca{O}(-1)\oplus\ca{O}(2) \Big)
{}~=~\ca{O}(-2)\oplus\ca{O}(-2)\oplus\ca{O}(1)$$
and we see that there are two zero modes for $\l^\z$ but none for $\l^\x$
or
for $\l^\eta$. Proceeding in this way we find the only zero modes
correspond
to the components:
 $$
\eqalign{
\ps_{\bar\x}  &={ \ba{\a}       \over (\ba{c}\ba{z}+\ba{d})   }~,\cr
\ps_{\bar\eta}&={ \ba{\b}        \over (\ba{c}\ba{z}+\ba{d})   }~,\cr
\ps^{\bar\z}  &={ \ba{\g}     \over (\ba{c}\ba{z}+\ba{d})   }
                +{ \ba{\d}/\ba{c}  \over (\ba{c}\ba{z}+\ba{d})^2 }~,\cr}
\hskip40pt
\eqalign{
\l_\x   &={{\a}\over cz+d}~,\cropen{3pt}
\l_\eta &={{\b}\over cz+d}~,\cropen{6pt}
\l^\z   &={{\g}\over cz+d}+
              {{\d}/c\over(cz+d)^2}~.\cr}
\eqlabel{instantini}
 $$
These results agree with those of\ \cite{{\DSWW,\DG}}.
\subsection{Vertex operators and powers of $g$}
The vertex operators for the {\bf27}-
and $\overline{\bf27}$-fields are conveniently written in terms of their
$SO(10)  \times U(1)\subset E_6$ content.
First, we list the vertex operators for the available
components of
\gen, \agen\ and the~\sing\ of $E_6$, all in the
ghost-number-$({-}1)$-picture and at zero
momentum~:
 $$
   V^I_{(-1)}~ = ~\ex{{-}\ph(\ba{z})}\>
                  h_{\ba\m}{}^\a(X)\>
                  \ps^{\ba\m}(\ba{z})\, \l^I(z)\, \l_\a(z)~,
\eqlabel{V10}
 $$
and
 $$
   V^0_{(-1)}~ = ~\ex{{-}\ph(\ba{z})}\>
                  \O_{\a\b\g}\, h_{\ba\m}{}^\a(X)\>
                  \ps^{\ba\m}(\ba{z})\, \l^\b(z)\, \l^\g(z)
\eqlabel{V0}
 $$
are the vertex operators for the (\ten,{\bf 1}) and the (\sing,{\bf -2})
component, respectively, of the \gen\ vertex operator.
The $\O_{\a\b\g}$ are the components of $\O$, the holomorphic three-form.
Similarly,
 $$
   V^{\bar J}_{(-1)}~ = ~\ex{{-}\ph(\ba{z})}\> b_{\a\ba\m}(X)\>
                  \ps^{\ba\m}(\ba{z})\, \l^{\bar J}(z)\, \l^\a(z)~,
\eqlabel{V10*}
 $$
and
 $$
   V^{\overline{0}}_{(-1)}~ = ~\ex{{-}\ph(\ba{z})}\>
                  \o^{\a\b\g}\, b_{\a\ba\m}(X)\>
                  \ps^{\ba\m}(\ba{z})\, \l_\b(z)\, \l_\g(z)
\eqlabel{V0*}
 $$
are the vertex operators for the (\ten,{\bf -1}) and the (\sing,{\bf 2})
component, respectively, of the \agen\ vertex operator.
We define $ \o^{\a\b\g}$ to be
$$  \o^{\a\b\g} \define {g^{\a\bar\l} g^{\b\bar\k} g^{\g\bar\s}
             \bar\O_{\bar\l\bar\k\bar\s}\over \|\O\|^2}
{}~~~\hbox{where}~~~\|\O\|^2 \define {1\over 3!}
g^{\k\bar\l}g^{\m\bar\n}g^{\r\bar\s}\O_{\k\m\r}\bar\O_{\bar\l\bar\n\bar\s}
{}~.\eqlabel{odef}$$
The (\sing,{\bf0}) vertex operator is
 $$
   V^1_{(-1)}~ = ~\ex{{-}\ph(\ba{z})}\>
                  s_{\ba\m}{}^\a{}_\b(X)\>
                  \ps^{\ba\m}(\ba{z})\, \l^\b(z)\, \l_\a(z)~.\eqlabel{V1}
 $$
Finally we record the ghost-number-$(0)$-picture dilaton vertex operator
 $$
V_R = i g_{\m\bar\n} (\partial x^\m)
\O^{\bar\n\bar\r\bar\s}\ps_{\bar\r}\ps_{\bar\s}~.\eqlabel{VR}$$
which comes from the pull-back of the \K\ form to the world--sheet.

The nonrenormalization theorem~\cite{\witten}~says that,
since the spacetime superpotential
must be a holomorphic function of the complex K\"ahler modulus, and
perturbative
corrections to the sigma model depend on the sigma model coupling constant
(the scale of the metric $g$), but not on the $\theta$-angle, there can be
no
perturbative sigma model corrections, either to the tree level, or to the
instanton  calculation of the spacetime superpotential.

Let us see how the counting of powers of the metric, $g$, goes, and
verify that, indeed, the
terms that appear in the spacetime superpotential  appear at order $g^0$.
We use the conventions in which the fermions have their indices raised and
lowered:
$\lambda^\m,\lambda_\n,\psi^{\bar\m},\psi_{\bar\n}$. In these conventions,
perturbation theory is not {\it manifestly\/} supersymmetric (as some of
the
indices on the fermions have been lowered), but this is more than
compensated
for by the fact that these conventions minimize the number of explicit
factors
of the metric which appear in the correlation functions. Since our
interest
will lie in calculating the spacetime superpotential, which receives no
perturbative corrections anyway, little will be lost by not using a
manifestly supersymmetric perturbation theory.

The forms $b_{\m\bar\n}(X)$, $h^\m{}_{\bar\n}(X)$, $s^\m{}_{\n\bar\r}(X)$
which
go into the definition of the vertex operators are taken to be
$g$-independent.
So, too, is $\O_{\m\n\r}(X)$, the holomorphic three-form.
The tensor $\o^{\m\n\r}$ defined in \eqref{odef} above
is also independent of $g$, as is clear from the relation
 $$
\O_{\m\k\l}\o^{\m\n\r} = \d_\k^\n \d_\l^\r - \d_\k^\r \d_\l^\n~.$$
However, $\O^{\bar\m\bar\n\bar\r}\asymp g^{-3}$, and
$\omega_{\bar\m\bar\n\bar\r}\asymp g^3$. This scaling with $g$ makes
sense, as
$\O_{\k\l\m}\o_{\bar\n\bar\r\bar\s}$ is the volume form on $M$, and so
scales like $g^3$.

When the fields have zero modes, one gets powers of $g$ in the path
integral
from doing the integration over the zero modes \
\ref{J.~Polchinski, \cmp{104} (1986) 37.}.
Each (complex) zero mode of $X$ introduces a factor of
$g$ into the path integral. Each zero mode of $\lambda^\m$ introduces a
factor
of $g^{-1/2}$ (and similarly for $\psi^{\bar\m}$), but each zero mode of
$\lambda_\n$ introduces a factor of $g^{1/2}$ (and similarly for
$\psi_{\bar\n}$), so the index theorem ensures that the {\it net} effect
of all
the fermi zero modes is to introduce no powers of $g$.
Each bose propagator brings down a power of $g^{-1}$, but the fermi
propagators
have {\it no} powers of $g$ associated to them.
Bringing down the four-fermi term,
$R^\m{}_\n{}^{\bar\r}{}_{\bar\s}\l_\m\l^\n\psi_{\bar\r}\psi^{\bar\s}$ from
the
action introduces a factor of $g^{-1}$, as that is how
$R^\m{}_\n{}^{\bar\r}{}_{\bar\s}$ scales.

With these conventions, one easily checks that all of the 3-point
functions we
have ever considered (both at tree level, and at the $n$-instanton level)
transform like $g^0$, as required.
For instance, the \yagen\  coupling at tree level has an explicit
$g^{-3}$ from the vertex operators, since two of the vertex operators
involve a
$b_{\m\bar\n}$ and are $g$-independent, but one, for the auxiliary field
in the
$({\bf 1,2})$ of $SO(10)\times U(1)$, has a
$b_{\m\bar\n}\o^{\m\k\l}\O^{\bar\n\bar\r\bar\s}$, and so
scales\Footnote{The spectral flow generator, which takes spacetime bosons
into
spacetime fermions (and spacetime fermions into the F-auxiliary field)
scales
like $g^{-3/2}$.} like~$g^{-3}$.
At tree level, there are 3 bose zero modes which leads to a factor of
$g^3$
from
the zero mode integrals, which yields
$g^{-3+3}=1$. At the $n$-instanton level (an $n$-fold cover of a line \
\Ref{\AspMor}{P.~Aspinwall and D. Morrison, \cmp{151} (1993) 245,
\hfill\break hep-th/9110048.}),
there are $2n+1$ bose zero modes and we need to bring down $2n-2$ factors
of
the four-fermi interaction from the actions, yielding
 $$
g^{-3+(2n+1)-(2n-2)}=1$$ as required.

The counting for the singlet couplings is
precisely analogous. The vertex operators in the correlation function
scale
as $g^{-3}$, as one of them is an $F$-auxiliary field (which carries a
factor
of $g^{-3}$), and the others are independent of $g$. No boson propagators
are
required, and the fermion dependence is such as to be able to absorb the
zero
modes present in an instanton background.
\subsection{Decomposition of singlets along a line}
Now a form
$ a^\m{}_\n=\End{a}{\m}{\n}d\bar\z \in H^1_{\delb}(\M,{\rm End}\>\T)$
decomposes into components that transform in line bundle
$\ca{O}(k)$ for $k=-3,0,3$,
 $$
\eqalign{
 a^\z{}_\z&\hskip60pt \ca{O}(0)\cr
 a^\z{}_j &\hskip60pt \ca{O}(3)\cr
 a^i{}_\z &\hskip60pt \ca{O}(-3)\cr
 a^i{}_j  &\hskip60pt \ca{O}(0)~.\cr}\eqlabel{decomp}
 $$
The decomposition of the tangent space to the manifold
at a point of the instanton
$$\T =\ca{O}(-1)\oplus\ca{O}(-1)\oplus\ca{O}(2)$$
is the statement that the normal (upper) indices
$\x$ and $\y$ count as ``charge''$-$1, while the
tangential index $\z$ has ``charge''$+$2.
Lower indices have the opposite charge. These cases are covered by another
special case of the Bott--Borel--Weil theorem, which states that
 $$
{\rm dim}H^1_{\delb}(S^2,\ca{O}(k))=\cases{0,  &for $k\ge-1$;\cropen{5pt}
                                        -k-1,~~&for $k\le -2$.\cr}
 $$
Thus the components of $ a^\m{}_\n$ are exact apart from the $ a^i{}_\z$,
so in \eq\eqref{singlets1} we may write
 $$\End{a}{i}{j}d\bar\z = \delb\a^i{}_j ~~~~\hbox{and}~~~~
   \End{a}{\z}{\z}d\bar\z = \delb\a~.\eqlabel{potentials}$$
We shall be concerned with these parts of the singlets in
Section~\chapref{yuk}
(the part $a^\z{}_j$ is also exact but plays no further r\^ole in our
analysis).
The components $a^j{}_\z$ represent a nontrivial cohomology group on the
instanton. It is natural to associate with $a^j{}_\z$ a (1,1)-form
 $$
a^j~\define~\End{a}{j}{\z}d\z d\bar\z$$
which for each value of $j$ takes values in $\ca{O}(-1)$. Alternatively we
may
regard $a^j$ as a \hbox{(0,1)-form} with values in
$\ca{O}(-1)\otimes\T^{\ast}_L=\ca{O}(-3)$, this being just a repetition
of
the statement in table~\eqref{decomp}. According to the Bott--Borel--Weil
Theorem, the cohomology group has dimension two. Stated differently: since
$a^j$ takes values in $\ca{O}(-1)$ there are two ways to integrate $a^j$
over
$L$ to get a number. There are two sections $(x^0,x^1)$ of $\ca{O}(1)$
over
$L$
which are the homogeneous coordinates of $L$ thought of as \cp1. The
product
$a^j x^\a$ takes values in the trivial bundle and so may be integrated
over
$L$. Taking $\z^\a = (1,\z)$ as local sections of $\ca{O}(1)$,
the burden of these remarks is that given a line $L$ together with an
$a^j$
the essential information is encoded in a $2\times 2$ matrix
 $$
A^{j\a}~\define~\int_L a^j \z^\a~. \eqlabel{integrals} $$

The matrix $A^{j\a}$ has indices of two distinct types. We are free to
redefine the coordinates $(\x,\eta)$
 $$
\pmatrix{\x\cr \eta\cr} \longrightarrow \pmatrix{\hat\x\cr \hat\eta\cr}~=~
\pmatrix{a&b\cr c&d\cr}\pmatrix{\x\cr \eta\cr}~.$$
The normal bundle is spanned by $d\x\wedge d\eta$ and this is unchanged if
the
matrix $\left( {a~b\atop c~d}\right)$ has unit determinant. Similarly we
may
redefine the basis $\z^\a$ by an independent $\hbox{SL}(2,\IC)$
transformation.
Under these transformations the matrix of $a^j$ transforms according to
the
rule
 $$
A^{j\a}\longrightarrow \hat{A}^{j\a}~=~U^j{}_k A^{k\b} V_\b{}^\a $$
with independent $\hbox{SL}(2,\IC)$ matrices acting on the left and on the
right.
Invariants may be formed by contracting indices with the permutation
symbols
$\ve_{jk}$ and $\ve_{\a\b}$. The invariant product of two matrices is
 $$
\ve_{jk} \ve_{\a\b} A^{j\a} B^{k\b}~.$$
{}From these simple observations, we conclude that each isolated line
can contribute to at most four eigenvalues of the mass matrix. Also,
much of the analysis of this article relevant to the Yukawa couplings
follows from the simple fact that there
is no invariant combination of three matrices.

\newpage
\section{mass}{Masses Generated by Instantons}
\vskip-20pt
\subsection{The mass matrix}
We wish to evaluate the correlation function corresponding to a
dilaton--singlet--singlet coupling $m(a,b) = \langle V_R V_a V_b \rangle$
where
$V_R$ corresponds to a dilaton  and $V_a$ and $V_b$  are vertex operators
corresponding to the singlets associated to forms $a^\m{}_\n$ and
$b^\m{}_\n$.
By referring back to Section 3.2 we see that we are to evaluate the
expression:
 $$
\eqalign{ \langle V_R V_a V_b \rangle = \mu\, \int d^6z d^4\a d^4\bar\a
& \Big[ \strut i g_{\m\bar\n} (\partial x^\m)
            \O^{\bar\n\bar\r\bar\s}
             \psi_{\bar\r}\psi_{\bar\s} \Big]_{(1)}\cr
&\times ~ \Big[ \strut \ex{{-}\ph}\>
                  a_{\bar\k}{}^\a{}_\b\>
                  \ps^{\bar\k}\, \l^\b\, \l_\a \Big]_{(2)}\cr
& ~~~~~~ \times ~ \Big[ \strut \ex{{-}\ph}\>
                  b_{\bar\l}{}^\g{}_\d\>
                  \ps^{\bar\l}\, \l^\d\, \l_\g \Big]_{(3)}~,\cr
}\eqlabel{RSS}$$
where we have written $V_R$, the vertex operator for the $F$--auxiliary
field,
in the 0--picture. The factor $\mu$, defined as
 $$
\m \define \left({\det(G^B) \over
\hbox{Det}'(\bar\partial^\dagger_\ca{T}
            \bar\partial^{\vphantom\dagger}_\ca{T})}\right)
\left({\hbox{Det}'(\bar\partial^\dagger_{\ca{T}\otimes\ca{S}}
        \bar\partial^{\vphantom\dagger}_{\ca{T}\otimes\ca{S}})\over
\det(G^F)}\right)~,\eqlabel{determinants}
 $$
represents the contributions from the integration over the
non--zero modes in the path integral and also the measure
of the parameter space of the zero modes.
The first factor is the contribution of bosonic
modes and the second factor is the contribution of fermionic modes.
Here, $G^B$ is the metric on the parameter space of bosonic zero modes and
$G^F$ the metric on the parameter space of fermionic zero modes.

We will now give a simple argument why $\m$ is constant in the basis
for the zero modes \eqref{instantini}. (For a more detailed discussion
of the values of the individual factors in~\eqref{determinants} see
Appendix~A.) This follows
from the Distler--Greene~\cite{\DG}
computation of the instanton contribution to the \yagen\
 Yukawa coupling since determinantal factors are the same in
each case (Distler and Greene use the same basis of fermionic zero modes
that
we have adopted here).
The vertex operators for the \yagen\ coupling
do not introduce any dependence on complex structure parameters and the
final
expression for the Yukawa coupling calculated by Distler and Greene does
not
depend on the complex structure parameters.
Indeed, if one calculates the ratio of the singlet mass and the
\yagen\ coupling, the determinants in question appear both in the
numerator and denominator and cancel out. Similarly, the right-moving
parts
of the correlation function are identical. Indeed, the only difference
between the two calculations occurs in the left-moving part of the
correlation function.

The evaluation of the zero mode integration in  expression \eqref{RSS}
is elementary given the table
\eqref{instantini} which records the nonzero components of the zero modes.
Thus the first expression in square
brackets in \eqref{RSS} simplifies to
 $$ \int d\ba{\b} d\ba{\a}~i g_{\z\bar\z}
{1\over{(c z_1 + d)^2}} \ve^{\bar\imath \bar\jmath}
\psi_{\bar\imath}(\ba{z}_1)\psi_{\bar\jmath}(\ba{z}_1) =
2 i g_{\z\bar\z} {1\over{|cz_1 + d|^4}}~.$$
{}From the second and third expressions we have a factor
 $$\eqalign{
\int d\ba{\d}\, d\ba{\g}~\psi^{\bar 3}(\ba{z}_2)\psi^{\bar 3}(\ba{z}_3) &=
\int d\ba{\d}\, d\ba{\g} \left( {\ba{\g} \over (\ba c \ba{z}_2{+}\ba d)} +
                         {\ba{\d}/\ba{c} \over (\ba c \ba{z}_2{+}\ba d)^2}
\right)
    \left({\ba{\g} \over (\ba c \ba{z}_3{+}\ba d)} +
      {\ba{\d}/\ba{c} \over (\ba c \ba{z}_3{+}\ba d)^2}\right)\cropen{7pt}
  &= {(\ba{z}_2-\ba{z}_3) \over (\ba c \ba{z}_2+\ba d)^2\, (\ba c
\ba{z}_3+
\ba d)^2}~,}
\eqlabel{QT}
 $$
which cancels the factor $(\ba{z}_2-\ba{z}_3)^{{-}1}$ from the free
correlator $\VEV{\ex{{-}\ph(\ba{z}_2)}\, \ex{{-}\ph(\ba{z}_3)}}$.

Proceeding in this manner we obtain an integral which is most simply
written
in terms of the variable $\z={az +b\over cz +d}$:
 $$\eqalign{
m(a,b) &\define \langle V_R V_a V_b \rangle =
\mu\,\int_L J ~\times~ \int_{L\times L}
      \ve_{jk} a^j (\z_2) b^k (\z_3) (\z_2 - \z_3) \cropen{5pt}
&= \mu\, \ve_{jk} \ve_{\a\b} A^{j\a} B^{k\b} \int_L J }
 \eqlabel{masses} $$
where $J = i g_{\z\bar\z} d\z d\bar\z$ is the \K\ form restricted to $L$.

Given a basis $a_I$ for $H^1(\M, \hbox{End}\>\T)$ and an instanton $L$ let
us
denote by the reduced mass matrix for $L$ the matrix
 $$
m^{(L)}_{IJ} ~=~ m^{(L)}(a_I,a_J) ~=~
\ve_{jk}\ve_{\a\b}\, A^{(L)\,j\a}_I A^{(L)\,k\b}_J ~,$$
with $A^{(L)\,j\a}_I$ the matrix of $a_I$ along $L$.
Note that, as expected from the discussion at the end of last Section,
$L$ can give mass to at most four singlets since among the matrices
$A^{(L)}_I$ there can be at most four that are linearly independent.

\subsection{Other splitting types}
The above analysis refers to instantons with normal bundle
\splittingtype{-1}{-1}. For some manifolds this is the generic situation.
For
a generic quintic hypersurface in $\cp4[5]$ there are 2875 lines
all of which have normal bundle \splittingtype{-1}{-1}. Other cases are
possible; the manifold may have instantons that form continuous families
and
for these instantons the normal bundle will be \splittingtype{p}{-p-2}\
for
some $p\geq 0$. It is known also, for example, that normal bundles of type
\splittingtype{0}{-2}\ and  \splittingtype{1}{-3}\ can arise in $\cp4[5]$
for
certain choices of defining polynomial. It is instructive to see how the
counting goes for these other cases.
\subsubsection{\splittingtype{0}{-2}}
For this case we take the normal coordinate $\x$ to be a local section of
$\ca{O}(0)$ and $\eta$ to be a local section of $\ca{O}(-2)$. By
considering
the transformation properties of the forms $a^\m=a^\m{}_\z d\z$ as in
Table~\eqref{decomp} we see that $a^\x$ and $a^\eta$ are nontrivial. The
form
$a^\x$ transforms in $\ca{O}(0)$ and has one degree of freedom while
$a^\eta$
transforms in $\ca{O}(-2)$ and has three degrees of freedom. In this case
we
again find a total of four degrees of freedom. These correspond to a
scalar
 $$
A~=~ \int_L a^\x$$
and a symmetric matrix
 $$
A^{\a\b}~=~ \int_L a^\eta \z^\a \z^\b~,~~~~\a,\b=0,1~.$$
Invariant bilinear products of two singlets on $L$ are for example
 $$
AB~~~~\hbox{and}~~~~\ve_{\a\g}\ve_{\b\d}\, A^{\a\b} B^{\g\d}~.$$
Note however that the coordinate $\x$ is not uniquely determined by the
statement that it is a section of $\ca{O}(0)$; since if $\x$ is such a
section then so is
 $$
\hat\x~=~\x + c_{\a\b}\z^\a \z^\b \y$$
for any coefficients $c_{\a\b}$. The change $\x\to\hat\x$ leads to a
change
in $A$
 $$
A~\to~A + c_{\a\b} A^{\a\b}~.$$
Since the contribution of a line to the mass matrix cannot depend on our
choice of coordinate $\x$ at first sight it would seem to
be the case that the contribution to
the mass matrix only involves the invariant quantity
$\ve_{\a\g}\ve_{\b\d}A^{\a\b}B^{\g\d}$.
In fact, from the fermionic zero-mode computation in section~4.1 one may
naively
conclude that the mass matrix is zero for an  \splittingtype{0}{-2} line.
As
will
be discussed in the next section and the appendix the
contribution is however non-zero. This is seen by studying the
degeneration,  parametrized by a parameter~$\e$, of lines of type
\splittingtype{-1}{-1}\
which coincide in the limit $\e\to 0$ to become an \splittingtype{0}{-2}\
line. The understanding of these contributions is
complicated by the existence of poles. We will find
that each individual line contributes a term to the mass matrix
which develops a  pole as $\e\to 0$ which is indeed of the form
$\ve_{\a\g}\ve_{\b\d}A^{\a\b}B^{\g\d}$.
These poles however cancel when summed over the lines.
There remains a finite contribution but this now involves further
invariants formed from derivatives of the~$A^{\a\b}$.
\subsubsection{\splittingtype{p}{-p-2} for $p\geq 1$}
For these cases we take the coordinates $\x$ and $\eta$ to be local
sections
of $\ca{O}(p)$ and $\ca{O}(-p-2)$. The form $a^\x$ is now trivial and we
are
thus left with $a^\eta$ which transforms in $\ca{O}(-p-2)$ and has $p+3$
degrees of freedom corresponding to a tensor
 $$
A^{\a_1\a_2\ldots\a_{p+2}}~=~ \int_L a^\eta
\z^{\a_1}\z^{\a_2}\cdots\z^{\a_{p+2}}~,~~~~\a_i=0,1~.$$
Note however that the invariant product between two of these
 $$
\ve_{\a_1\b_1}\ve_{\a_2\b_2}\ldots\ve_{\a_{p+2}\b_{p+2}}
{}~A^{\a_1\a_2\ldots\a_{p+2}}B^{\b_1\b_2\ldots\b_{p+2}}$$
is symmetric for $p$ even but antisymmetric for $p$ odd so for instantons
corresponding to odd $p$ these terms
cannot contribute to the mass matrix.
\newpage
\REF{\rkatz}{S. Katz, Compos. Math. {\bf 60} (1986) 151.}
\REF{\rharris}{J. Harris, Duke. Math. J. {\bf 46} (1979) 685.}
\section{quintic}{Examples of the Effects of Lines in $\cp4[5]$}
\vskip-20pt
\subsection{General facts}
Given a line, $L$, we may choose coordinates adapted to the line such that
$L$
corresponds to the equations $X_2=X_3=X_4=0$. In terms of these
coordinates the
quintic, $p$, that defines the \cym\ takes the form
 $$
 p = X_2 F + X_3 G + X_4 H + K~,$$
$F$, $G$ and $H$ being quartics in $X_1$ and $X_5$,
and $K$ being a quintic of order 2 or more in $X_2$, $X_3$ and $X_4$. The
splitting type of the normal bundle of $L$ depends on the degree of
independence
of the quartics $F$, $G$ and $H$. Katz \cite{\rkatz} has shown that if
there
are
no linear polynomials $\ell_1(X)$, $\ell_2(X)$, $\ell_3(X)$ such that
 $$
\ell_1(X) F + \ell_2(X) G + \ell_3(X) H = 0~,$$
such a relation being termed a quintic relation, then the normal bundle is
\splittingtype{-1}{-1}. If there is a quintic relation but no quartic
relation
 $$
c_1 F + c_2 G + c_3 H = 0~,$$
with constants $c_1$, $c_2$, $c_3$, then the normal bundle is of type
\splittingtype{0}{-2}. The final case is if there is a quartic relation
(\ie  $F$, $G$ and $H$ are not linearly independent) then the normal
bundle is of type \splittingtype{1}{-3}. For $\cp4[5]$ no other case is
possible
\ \cite{\rharris} \
in virtue of the fact that the normal bundle of the line in \cp4[5] is
a subbundle of $L$ when thought of as a line in \cp4.
Since the normal bundle of the line in \cp4 is $\ca{O}(1)^3$,
the normal bundle in \cp4[5],  \splittingtype{e_1}{e_2},
must have both $e_1,e_2 \leq 1$.

Examples of such functions are
 $$
F=X_1^2X_5^2~~~,~~~G=X_5^4~~~,~~~H=X_1^4$$
for which there is no quintic relation and
 $$
F=X_1^3X_5^{\hphantom{2}}~~~,~~~G=X_5^4~~~,~~~H=X_1^4$$
for which there is a quintic relation but no quartic relation.
We shall be concerned, in the following, with a degenerating family for
which
 $$
F=X_1^3X_5 + \e X_1^2X_5^2~~~,~~~G=X_5^4~~~,~~~H=X_1^4$$
such a line has normal bundle \splittingtype{-1}{-1}\ for each nonzero
$\e$ but
the normal bundle is \splittingtype{0}{-2}\ when~$\e=0$.

In order to find the normal coordinates $\x$, $\eta$, we can follow the
prescription of Harris~\cite{\rharris}.
We first look for sets of functions of weight one in $X_1$ and $X_5$,
${\bf m}=(m_2,m_3,m_4)$, such that
 $$
m_2 F + m_3 G + m_4 H =0~.\eqlabel{mdef}$$
Each ${\bf m}$ will be used to build a normal coordinate.
For a coordinate in an $\ca{O}(-n)$ bundle,
${\bf m}$ should be defined in the patch $U_1$ (\ie when $X_1\neq 0$), and
$\left({X_1\over X_5}\right)^n {\bf m}$ must be defined in $U_5$.
These are then used to define the normal vectors in the appropriate
patches.
Since we have two normal coordinates, there
are two sets of functions, ${\bf m}$ and ${\bf m'}$.
Let $(\x,\eta)$ be the normal coordinates over $U_1$ and
$(\tilde\x,\tilde\eta)$ the normal coordinates over $U_5$. On $U_1$ we set
 $$\eqalign{
\pd{}{\x} &= m_2 \pd{}{X_2} + m_3 \pd{}{X_3} + m_4 \pd{}{X_4} \cr
\pd{}{\y} &= m'_2 \pd{}{X_2} + m'_3 \pd{}{X_3} + m'_4 \pd{}{X_4} \cr}
\eqlabel{vecdef}$$
and on the intersection $U_1\cap U_5$ we find
$\pd{}{\tilde \x} = \left({X_1\over X_5}\right)^n \pd{}{\x}$
and $\pd{}{\tilde \y} = \left({X_1\over X_5}\right)^{n'} \pd{}{\y}$
for an $\ca{O}(-n)\oplus\ca{O}(-n')$ bundle.

It is perhaps worth recording here also that since the only splitting
types for
$\cp4[5]$ are \splittingtype{-1}{-1}, \splittingtype{0}{-2}\ and
\splittingtype{1}{-3} the data describing the restriction of the singlet
to a
line consists in each case of four numbers. These are the quantities
$A^{j\a}$,
$A$ and $A^{\a\b}$, and $A^{\a\b\g}$ for the three cases respectively. We
can
understand this by noting that a singlet $\ssa_M dx^M$ when restricted to
the
line $X_2=X_3=X_4=0$ takes the form
 $$
\ssa_MdX^M = (\ssA_0 X_1^3 + \ssA_1 X_1^2X_5 + \ssA_2 X_1X_5^2 + \ssA_3
X_5^3)
(X_1dX_5 - X_5dX_1) \eqlabel{singrep}$$
which is specified by giving the four coefficients $\ssA_0,\ldots,\ssA_3$.

A remarkable fact demonstrated by Harris~\cite{\rharris} is that an
arbitrary permutation of the lines may be achieved by monodromy. That is,
if
the manifold is deformed around a closed loop in the moduli space then the
manifold returns to the original manifold but with the lines permuted and
an
arbitrary permutation can be achieved by suitable choice of loop. Harris
demonstrates this by showing that there are monodromies that interchange
any
two given lines while leaving the remaining lines unchanged.

For the case of $\cp4[5]$, we may choose a basis for the singlets
that is independent of the complex structure. We may do this by forming
from
the set of all vectors $\hbox{\ss s}_A(X)$ of quartic monomials (of which
there are $5\times 70=350$) fixed linear combinations that satisfy
$X^A \hbox{\ss s}_A(X)=0$.
We are left with a basis of 224 vectors of quartics with fixed
coefficients.
In this section we will first study the mass
matrix
corresponding to an \splittingtype{-1}{-1}\ line. We will then study how
the
matrix
behaves as three \splittingtype{-1}{-1}\ lines come together to form an
\splittingtype{0}{-2}\ line. To this end we examine first the geometry of
the
normal
bundle of an \splittingtype{0}{-2}\ line and then we study the
degeneration.

\subsection{An \splittingtype{0}{-2} line as the limit of
            \splittingtype{-1}{-1} lines}
Consider the pair of lines in \cp4 given parametrically by
 $$
l_{\pm\e} = (u,~\pm\e u,~0,~\mp\e v,~v)~,   \eqlabel{twolines}$$
where $(u,v)$ are the homogeneous coordinates of the \cp1.
These are described by the equations
$$\eqalign{
\e^2 X_1^2 - X_2^2 &=0 \cr
\e^2 X_5^2 - X_4^2 &=0 \cr
X_1 X_4 + X_2 X_5  &=0 \cr
\e^2 X_1 X_5 + X_2 X_4 &=0 \cr
X_3 &=0 ~,\cr}$$
and so are embedded in any
hypersurface $p=0$ in \cp4[5] with polynomial
$$ p = (\e^2 X_1^2 - X_2^2) Q + (\e^2 X_5^2 - X_4^2) \tilde Q
       + (X_1 X_4 + X_2 X_5) S + (\e^2 X_1 X_5 + X_2 X_4) R + X_3 G$$
with $Q$, $\tilde Q$, $S$ and $R$ cubics, $G$ a quartic.

Firstly note that when $\e=0$, the lines coincide, and are given by
$X_2=X_3=X_4=0$.
Linearizing in $X_2$, $X_3$ and $X_4$, the polynomial $p$ becomes
$$ p = X_2 X_5 S + X_3 G + X_4 X_1 S~,$$
and so this line has normal bundle
$\ca{O}(0)\oplus\ca{O}(-2)$
for sufficiently general $S$ and $G$.
We will choose
$$      S|{}_{l_{\pm\e}} = u^3 ~,~
        G|{}_{l_{\pm\e}} = v^4 ~.$$

If $p$ is sufficiently general then for $\e\neq 0$ the lines have normal
bundle
$\ca{O}(-1)\oplus\ca{O}(-1)$.
Since the line $l_\e$ is defined as $Y_2 = Y_3 = Y_4 = 0$
for the variables
$$\eqalign{
   Y_2 &= X_2 - \e X_1 \cr
   Y_3 &= X_3   \cr
   Y_4 &= X_4 + \e X_5 \cr} $$
we linearize the polynomial in the $Y$'s to find
$$p|_{l_{\e}} = Y_2 F + Y_3 G + Y_4 H $$
with the quartic functions
$$ F = X_5 S - 2\e X_1 Q -\e X_5 R\quad,\quad
   H = X_1 S + 2\e X_5 \tilde Q + \e X_1 R  ~.$$

For definiteness, we choose a polynomial with
$$\eqalign{
        Q &= - {1\over 2} X_1 X_5^2 - {1\over 2} \k X_5^3
                                            + \Delta Q ~~~,\cr
 \tilde Q &= - X_1 T  ~~~,\cr
        S &= X_1^3 - X_4 T - {1\over 4} (\e^2 X_1 X_5 + X_2 X_4) X_5
                                            + \Delta S ~~~,\cr
        R &= X_5 T ~~~,\cr
        G &= X_5^4 +  \Delta G~,\cr}$$
where T is a quadratic polynomial and $\Delta Q$, $\Delta S$ and
$\Delta G$ are generic polynomials of the appropriate degree which are
are chosen to vanish on both lines $l_{\pm\e}$.
$\Delta Q$, $\Delta S$ and $\Delta G$ are  introduced only to ensure the
transversality of $p$ and are otherwise irrelevant in the discussion
below.

The linearizations of $p$ about the lines $l_{\pm\e}$ are particularly
simple.
The linearization about $l_{\e}$ takes the form
 $$
p|_{l_{\e}} =
  Y_2 \,(X_1^3 X_5 +\e X_1^2 X_5^2 + \k\e X_1 X_5^3) +
                                        Y_3 \,X_5^4 + Y_4 \,X_1^4~,
   \eqlabel{peps}$$
in which we see clearly that the normal bundle is \splittingtype{-1}{-1}
for $\e \neq 0$ and is \splittingtype{0}{-2} when $\e=0$.

The calculation of the contribution of each line to the mass matrix
$\e_{ij}\e_{\a\b} A^{i\a} B^{j\b}$ is
straightforward. We need to calculate the matrices $A^{i\a}$ given by
$$ A^{i\a} = \int_L\, a^j \z^\a ~, $$
where we use the representation for the singlets~\cite{\rPhilip}\ as
discussed in section~2:
$$
a^\mu{}_\nu =
 -{1\over 2\p i}\,\ssa^\a{}_\nu\ \ch^\mu{}_{\bar\rho\a}\ \rd X^{\bar\rho}
            = -{1\over 2\p i}\,\ssa^\a{}_\nu\,
h_{\a\bar\b}\,\ch_{\bar\t\bar\rho}{}^{\bar\b} \,
     g^{\mu\bar\t}\,\rd X^{\bar\rho}~.
\eqlabel{sings} $$

We need to find first the normal coordinates $(\xi, \, \eta)$, which
as outlined in the previous subsection, is done by solving for
${\bf m}=(m_2,m_3,m_4)$ in $m_2 F + m_3 G + m_4 H =0$.  Since the
lines $l_{\pm\e}$ are \splittingtype{-1}{-1} for
$\e\ne\ 0$, the $m$'s are all quadratic in $\z$.  We find then that in
the coordinate patch with $X_1\ne 0$ we have the following relations:
\vskip10pt
$$
\hbox{\fourteenit l}_{+\e}~~~~ \left\{~~~\quad
\eqalign{
{Y_2\over X_1} &~=~ ( - 1 + \k\z)\x +
           {1\over {\e(\e - \k)}} ( 1 - \e\z )\z\,\y \cropen{5pt}
{Y_3\over X_1} &~=~ - \k^2\e\,\x +
         \left( 1 + {{\k\e}\over{\e-\k}}\z \right)\y \cropen{5pt}
{Y_4\over X_1} &~=~ (1 + (\e-\k)\z )\z\,\x
                     - {1\over {\e(\e - \k)}}\z^2\,\y \cropen{5pt}
{X_5\over X_1} &~=~\z \cropen{5pt}    }\right.
\eqlabel{minusoneminusone}$$

We work with the Fubini--Study metric on \cp4
which descends to a metric in the neighbourhood of the instanton as
 $$
g_{i\jbar} \rd X^i \rd X^{\jbar} = \left(
  {\d_{i\jbar}\over \s}- {X_i X_{\jbar} \over \s^2}\right)\rd X^i
\rd X^{\jbar}~, \eqlabel{instmetric}$$
where $i$ and $\jbar$ run over the values for the inhomogeneous
coordinates,
and $\s = \sum_{A} |X^A|^2$. The components of interest of
the extrinsic curvature are calculated from~\cite{\rPhilip}
$$\chi_{\mu\nu}{}^\a = \pd{X^A}{x^\mu}\pd{X^B}{x^\nu}
                   \ppd{p^\a}{X^A}{X^B}~,$$
as discussed in Section 2.1.
For a polynomial $p$, the factor $h_{\a\bar\b}$
(the index $\a$ takes only one value
in our case and we now drop this index for notational simplicity) is given
by
$$
h^{-1}= \s \sum_A |\partial p/\partial X^A|^2 ~.$$
For the singlets that are nonzero along the instanton, the factor
$\ssa_\nu$
has the form $\ssa_\z= X_1^4 \, \ssA(\z)$, with
$\ssA(\z) = \sum_{i=0}^3 \ssA_i \z^i$,
which finds its origin in the assignment \eqref{singrep}.

For the line $l_\e$ the only relevant components of the
extrinsic curvature are
$$\eqalign{
    \chi_{\x\z} &= X_1^5 [ - 1 + (- 2\e + \k)\z - \k\e\z^2 - \k^2\e\z^3]
\cropen{3pt}
    \chi_{\y\z} &= X_1^5 {\z\over {\e(\e -\k)}}
                         [ 1 + \e\z + (2\e-\k)\e\z^2 +\k\e^2
\z^3]~~.\cr}$$
Putting all the information together we obtain the matrix for the singlets
$$
A^{j\a}(l_\e) =  \pmatrix{
\bigfract{\ssA_2}{\e(\e-\k)} - \bigfract{\ssA_1}{\vphantom{\e}(\e-\k)}&
\bigfract{\ssA_1}{\e(\e-\k)} - \bigfract{\ssA_0}{\vphantom{\e}(\e-\k)}
\cropen{8pt}
 \ssA_3 - \k\ssA_2 & \ssA_2 - \k\ssA_1\cr}~.$$
The surprising conclusion is that $l_\e$ contributes a term to the mass
matrix
that has  a pole as the lines coalesce at $\e = 0$
 $$
  m~\asymp~ {1\over {\e(\e - \k)}} (\ssA_2^2 - \ssA_1\ssA_3) ~~.
                \eqlabel{mpole}$$
However, when the contribution of the pair of lines
$l_{+\e}$ and $l_{-\e}$ is added up, the pole at $\e = 0$ cancels except
when $\k=0$.  In the case of $\k=0$, the effect as $\e\to 0$ is that
the leading term of order $1/\e^2$ is multiplied by two and
the terms of $\ca{O}(\e^{-1})$ cancel.

What is happening is that the polynomial $p$ has a third line
that when $\k=0$ also coalesces with the lines $l_{\pm\e}$ as
$\e\to 0$.  For a certain choice of the polynomials $\Delta Q$,
$\Delta S$ and $\Delta G$,  this line  is given by
$$
   l_{\e^2} = \left(u,~{1\over 2}\e^2 v,~0,~0,~v\right)~.
                   \eqlabel{thirdline}$$
The situation in this case is very interesting since now
there are three lines that coalesce at $\e = 0$.
The line $l_{\e^2}$ is defined as $Z_2 = Z_3 = Z_4 = 0$ for
the variables
$$\eqalign{
   Z_2 &= X_2 - {1\over 2} \e^2 X_5 \cr
   Z_3 &= X_3   \cropen{3pt}
Z_4 &= X_4~~.   \cr} $$
Linearizing the polynomial in the $Z$'s we find
$$
p|_{l_{\e^2}} =
  Z_2 \,\left(X_1^3 X_5 + {1\over 4} \e^2 X_1 X_5^3\right) +
  Z_3 \,X_5^4 +
        Z_4 \,\left(X_1^4 - {1\over 4} \e^2 X_2 X_5^2 -
                       {1\over 16} \e^4 X_5^4\right)   ~.
   \eqlabel{peps2}$$
from which we see that $l_{\e^2}$ is indeed of type
\splittingtype{-1}{-1}.

Following again the procedure above to find the contribution to the mass
matrix for this line, we find for the normal coordinates $(\x,\,\y)$
 $$
\hbox{\fourteenit l}_{\e^2}~~~~ \left\{~~~\quad
\eqalign{
{Z_2\over X_1} &~=~\left( -1 + {\e^2\over 2}\z^2\right)\x
          - {2\over\e^2}\z\,\y\cropen{5pt}
{Z_3\over X_1} &~=~ - {\e^4\over 16}\z\,\x
          + \left( 1 + {\e^2\over 8}\z^2\right)\y\cropen{5pt}
{Z_4\over X_1} &~=~ \z\,\x + {2\over\e^2}\z^2\,\y\cropen{5pt}
{X_5\over X_1} &~=~\z~. \cropen{5pt}     }\right.
\eqlabel{minusoneminusone2}$$
\vskip10pt
\noindent
The only nonzero components of the extrinsic curvature are now
$$\eqalign{
        \chi_{\x\z} &= - X_1^5 {1\over 8}
            \left(8 + 6\e^2\z^2 +\e^4\z^4 \right)\cropen{3pt}
        \chi_{\y\z} &= - X_1^5 {\z\over {2\e^2}}
            \left(4 - 3 \e^2\z^2 \right)~~.\cr}$$
The coefficient matrix for the singlets is now
$$
A^{j\a}(l_{\e^2}) =  \pmatrix{
-2 \bigfract{\ssA_2}{\e^2} - \bigfract{\ssA_0}{2\vphantom{{}^2}}&
-2 \bigfract{\ssA_1}{\e^2} \cropen{8pt}
  \ssA_3 - \bigfract{\e^2 \ssA_1}{4}&
  \ssA_2 - \bigfract{\e^2 \ssA_0}{4}\cr}~,$$
from which it is obvious that its contributions to the mass
matrix will be such that they precisely cancel the pole term
in \eqref{mpole}\ for $\k=0$.
One can easily check that the  mass matrix obtained after summing
over the three lines is non--vanishing.  The total mass matrix
however is obtained after summing over {\it all} the lines in the
manifold.
and this is not practical with the methods presented here.
\subsection{The singlet mass matrix is probably always zero}
There is a general argument, rather similar to the one in
Ref.~\cite{\SW},
to the effect that the superpotential should in fact vanish, and with it
the
mass matrix for the {\bf1}'s. The present context being somewhat
different, we
wish to present this argument and also to discuss its possible
limitations.

The superpotential in the effective 4-dimensional theory is a
complex-analytic section of a holomorphic line-bundle over the space of
complex structures. Then, if it is singular, this must happen on a
subspace of
codimension~1 in the moduli space. Also, the superpotential cannot
possibly be
singular where the \cy\ manifold is smooth. Therefore, the subset of the
moduli space where the superpotential may be singular must be open and
dense
in the discriminant locus\Footnote{The discriminant locus is the subspace
in
the space of complex structures of a complex manifold where the manifold
itself is singular.}---which we will prove is impossible at least for a
great
majority of \cy\ manifolds. Finally, being a non-singular analytic section
of
a non-trivial holomorphic bundle over a compact moduli space, it in fact
has
to be zero. For technical reasons explained below, we will consider a
somewhat
redundant but non-singular moduli space.

For the sake of clarity, we first discuss the case of the quintic in
$\cp4$ and
turn to generalizations below. The effective moduli space of all quintics
(singular ones included) is rather badly singular and many of the standard
genericity arguments do not apply straightforwardly. For this reason, we
shall
instead formulate the argument in the (projective) space of
coefficients.  Since there are
126 quintic monomials on $\cp4$, the projective space of coefficientsis
$\cp{125}$; the true (effective) moduli space is obtained by passing to
the
$PGL(5;\IC)$ quotient. It is standard that the subspace of singular
quintics
is of codimension~1 in $\cp{125}$, and its $PGL(5;\IC)$ quotient is the
true
discriminant locus (see section 2.2.1 of Ref.~\cite{\rBeast} for details).
 Also, the generic point in this 124-dimensional subspace parametrizes a
quintic with a single node\Footnote{For an explicit example of a quintic
with a single node, see %
Ref.~~\Ref{\rRolling}{P.~Candelas, P.~Green and T.~H\"ubsch,
\npb{330}(1990)49--102}, or section D.3.3 of Ref.~\cite{\rBeast}.
}, for which the defining equation must take the
following form
$$ p(x) = (x_1x_4-x_2x_3) + p_3(x) + p_4(x) + p_5(x)~,
\eqlabel{nodalP}$$
where we chose coordinates so that the node occurs at the point
$(0,0,0,0,1)$
in the coordinate patch with $x_5=1$, and where the $p_k(x)$ are
polynomials
of order $k$ in $x_1,\ldots,x_4$.

Now, for a line parametrized by $\l$ which passes through the node
$(0,0,0,0,1)$, the local neighborhood of the node may be parametrized as
$(\l s_1 t_1,\l s_1 t_2,\l s_2 t_1,\l s_2 t_2,1)$. The coordinates
$(s_1,s_2;t_1,t_2)$ are homogeneous coordinates of $\cp1{\times}\cp1$, the
local neighborhood of the node projectivised along $\l$. It is easy to see
that the quadratic part of~\eqref{nodalP} vanishes with this
parametrization.
The remaining terms $p_3,p_4,p_5$ must vanish separately, as they appear
with
different powers of $\l$. Thus, they impose three independent conditions
on
$\cp1{\times}\cp1$, and so have no common solution in general. However,
let
some $\tilde s, \tilde t$ be a solution of, say, $p_3=0=p_4$. The
remaining
equation $p_5(\tilde s,\tilde t)=0$ may then be considered as a single
condition on the parameters, and so all three $p_3,p_4,p_5$ vanish on a
codimension~1 subset of the 124-dimensional parameter space of singular
quintics. This codimension~2 subset of the projective parameter space
$\cp{125}$ is then the largest subset of the parameter space over which
the
corresponding \cy\ manifold contains at least a simple node {\it and} a
line
passing through it, and is also the largest subset where the
superpotential
may possibly diverge.
 However, this is manifestly neither open nor dense in the codimension~1
discriminant locus, and we conclude that the effective 4-dimensional
superpotential cannot diverge. Finally, being an everywhere finite
analytic
section of a non-trivial holomorphic line-bundle over the projective
parameter
space $\cp{125}$, the superpotential must in fact vanish. So must
also the mass-matrix and all the Yukawa couplings which involve the
{\bf1}'s to all orders.

It is possible that this argument does not generalize readily to
other known manifolds, and we now discuss to some possible limitations.
Firstly, there is the opposite extreme of the above situation---by now
familiar from the study of the mirror of $\cp4[5]$~
(see Ref.~~\Ref{\rCdGP}{P.~Candelas, X.~de~la~Ossa, P.~Green and
L.~Parkes,
\npb{359}(1991)21.}).\
Here, the space of true (effective) complex structures is 1-dimensional,
and
the above genericity arguments may be thwarted simply by too low total
dimension. Similar limitations will occur also in other cases with
low-dimensional moduli spaces.

Recall now that the 1-parameter family of mirrors of $\cp4[5]$ is
constructed
as a $\IZ_5^{~3}$-orbifold of a 1-parameter family of
$\IZ_5^{~3}$-symmetric
quintics. The difficulties in this case may then also be understood as a
consequence of the explicit restriction to highly non-generic
quintics. In fact we will work through the case of the mirror quintic
below. The
above general argument may then appear similarly thwarted in cases where
the
considered set of \cy\ manifolds is restricted to have special symmetries,
and
such cases must be re-examined case by case. Whether or not the argument
can
always be made by deforming into (0,2) theories remains unclear at this
time.

Finally, there may well exist \cy\ manifolds for which the discriminant
locus
is still of codimension~1 in the space of complex structures, but the
mildest
possible singularization is worse than a conifold with a single node. For
example, if it should happen that the mildest singularization includes a
singular curve, the world sheet instanton might map the entire world sheet
into the singular curve, and our above argument would fail. While we were
not
able to find an explicit counter-example of this kind, it does not
seem  impossible that
manifolds embedded in weighted projective spaces might exhibit this
phenomenon
owing to special (in)divisibility properties of the weights. Also,
examples of
\cy\ manifolds exist~(see section 3.5.1 of Ref.~\cite{\rBeast}) which are
defined by
overdetermined systems of equations, and about which far too little is
known
to make any definite claims.

However, one general comment is in order for these latter cases, and in
fact
for all possible candidate counter-examples. Namely, it is well known that
not
all deformations of the complex structure of an abstract \cy\ manifold
need
be representable as deformations of a given, concrete embedding. It is
then
perfectly possible that the considered embedding is constrained in just
such a
way that the mildest singularization of the embedding develops a singular
curve, rather than an isolated node. This by itself would {\it not}
constitute
a counter-example to the above general argument, as long as a deformation
of
the abstract manifold exists which smoothes the singular curve(s) into one
or
more isolated nodes. Unfortunately, effects of such (abstract)
deformations of
the manifold are much harder to study than the explicitly realizable
(polynomial) embedding deformations. Therefore, the rigorously proven {\it
full} scope of the above claim, that all the terms in the
superpotential which involves should vanish, remains an open question.

Silverstein and Witten~\cite{\SW}\ have stressed the role of the mass
matrix  as an
obstruction to the possibility of deforming a $(2,2)$-theory to a $(0,2)$
theory. We have argued above that, in simple cases, the total mass is zero
since for a generic conifold the instantons do not pass through the node.
For
particular manifolds however this may not be true. An example of this is
the
mirror of
$\cp4[5]$.
This manifold~~\Ref{\GP}{B.~R.~Greene and M.~R.~Plesser,
\npb{338}(1990)15},  which we shall call $\ca{W}$, can be
realized by identifying the points of a covering manifold $\widehat\ca{W}$
that
corresponds to the quintic
 $$
p \define X_1^5 + X_2^5 + X_3^5 + X_4^5 + X_5^5 - 5\ps\,X_1X_2X_3X_4X_5 =
0
\eqlabel{mirrorpoly} $$
in $\cp4[5]$ under the action of a group which is abstractly $\IZ_5^3$ and
which has the generators
 $$\eqalign{
&(\IZ_5;\,1,0,0,0,4)\cr
&(\IZ_5;\,0,1,0,0,4)\cr
&(\IZ_5;\,0,0,1,0,4)~.\cr}$$
The notation indicates that the first generator, for example, has the
action
 $$
(X_1,X_2,X_3,X_4,X_5) \to (\a X_1,X_2,X_3,X_4,\a^4 X_5)~;
{}~~~~\a=e^{{2\p i\over 5}}~.$$
Now the polynomial \eqref{mirrorpoly} is constrained to have only one
parameter by
the symmetries. The manifold $\widehat\ca{W}$ becomes singular when
$\ps^5=1$. The
resulting variety was studied by Schoen~~\Ref{\CS}{C.~Schoen, J. f\"ur
die Reine und Angewandte Math. {\bf 364}(1986)201.}\ because of its
special  properties before
its importance for mirror symmetry was realized. Before taking the
quotient by
the group the variety $\widehat\ca{W}$ has 125 nodes at the points
 $$(\a^{n_1}, \a^{n_2}, \a^{n_3}, \a^{n_4}, \a^{n_5})\qquad\hbox{with}
\qquad \sum_i\,\a^{n_i} = 0~.$$
These points are all identified under the group so that $\ca{W}$ has only
one
node. All the nodes are equivalent in virtue of the symmetries so consider
for
example the node $(1,1,1,1,1)$. Schoen has observed that the 24 lines that
connect this node to the others whose coordinates are the permutations of
$(\a,\a^2,\a^3,\a^4,1)$ actually lie in the hypersurface $p=0$. The group
identifies the nodes and the lines so that on $\ca{W}$ we not only
have one  line
that passes through the node but the line also has self intersection at
the
node!

In this situation we can have recourse to an argument of Silverstein and
Witten and deform the theory into a $(0,2)$ theory. In the context of the
linear sigma model~~\Ref{\Phases}{E. Witten, \npb{403} (1993) 159.}\ the
instanton contribution can only be singular if an instanton passes
through a point $X^\sharp$ for which
 $$
\pd{p}{X_i} + q_i = 0 \eqlabel{Xsharp}$$
with $q_i$ a vector of quartics corresponding to a singlet deformation. It
now
suffices merely to show that no instanton passes through the point
$X^\sharp$
specified by this equation. On $\ca{W}$ there are four quartic forms
that are not derivatives of quintics. Let
 $$
r~=~X_1X_2X_3X_4X_5\qquad\hbox{and}\qquad r_i~=~{r\over X_i}$$
so that $r_i$ is the product of the $X_j$ with $X_i$ deleted. There are
five
forms $r_i\,dX_i$ that are invariant under the group but there is a
relation
 $$
\sum_i\,r_i\,dX_i~=~dr$$
so only four are independent. This being so we take for the singlet
deformation
 $$
q~=~5\e\sum_{i=1}^5\,\l_i r_i dX_i\qquad\hbox{with}\qquad
\sum_{i=1}^5\,\l_i~=~0~.$$
The equations $p_{,i} + q_i = 0$ can be solved only when
  $$
\prod_{i=1}^5\, (\ps - \e\l_i)~=~1 \qquad\hbox{and}\qquad
X_i~=~(\ps - \e\l_i)^{1\over 5}~,$$
where, by means of the scaling symmetry, we have set $r=1$ (note that $r$
cannot be zero or all the $X_i$'s would vanish too).
We can, for simplicity, solve these equations perturbatively in $\e$. We
find
 $$
\ps ~=~ 1 - \e^2\sum_{i>j}\, \l_i\l_j\qquad\hbox{and}\qquad
X_i ~=~ 1 - {\e\over 5}\l_i~.$$
If we impose the constraint
 $$
\sum_i\,\l_i^2~=~0 \qquad\hbox{in addition to}\qquad \sum_i\,\l_i~=~0$$
then we fix $\ps = 1$ and we are able to move $X^\sharp$ in a three
dimensional
neighborhood of the node. Since there are points in such a neighborhood
through
which lines do not pass it is clear that the one-instanton contribution to
the
mass cannot be singular and hence must vanish. The conclusion is that
the total mass and the Yukawa couplings that involve \sing's vanish in
this case. Note however that the argument turned out to be rather
delicate. A self intersecting line always passes through the node and
of the 224 \sing's (there must be 224 by mirror symmetry) only four
can be represented by polynomials. These four however are sufficient
to complete the argument.

\newpage
\section{yuk}{Yukawa Couplings}
\smallskip
\noindent
We now turn to evaluating the \ymix\ and then \ysing\ couplings using the
general formula~\eqref{genyukawa}.
\subsection{The coupling in terms of zero modes}
We have chosen local coordinates $X^1,\, X^2$ and $X^3$ so that
$X^3=\z$ is aligned with the instanton as in \eqref{inst.eq}.
The instantino modes are given in Eq.~\eqref{instantini}.
Note that we have used the components of the metric tensor to lower the
indices; in fact, that is how they will naturally occur in the vertex
operators in Eq.~\eqref{V10}--\eqref{V1}.

Before turning to the full calculation, notice that the couplings
that interest us have a common factor.   Consider first the integral
over the $\bar\psi$'s.
Since the ``0-picture'' of the vertex operators is obtained by
replacing the
$\ex{{-}\ph(\ba{z})}\, \ps^{\ba\m}(\ba{z})$ field combination with
$\O^{\ba\m}{}_{\n\r}\, \ps^\n(\ba{z})\, \ps^\r(\ba{z})$
and since there are two $\ps^{\bar3}$-instantinos, we always
obtain the factor
 $$
  \int \rd\ba{\d}\, \rd\ba{\g}~\ps^{\bar\z}(\ba{z})\ps^{\bar\z}(\ba{w})~
   = ~{(\ba{z}-\ba{w}) \over (\ba{c}\ba{z}+\ba{d})^2\,
(\ba{c}\ba{w}+\ba{d})^2}~,
 $$
which, as in our discussion of the mass matrix, again cancels the factor
$(\ba{z}-\ba{w})^{{-}1}$ resulting from the free correlator
$\VEV{\ex{{-}\ph(\ba{z})}\,
\ex{{-}\ph(\ba{w})}}$. Moreover, the two $\bar\ps$-instantinos
necessarily occur on different vertices, while $\ps\,\ps$ occurs on the
third vertex, so no selection rule can ever come from the
$\ps$-integration.

On the other hand, the $\l$'s and the $\bar\l$'s are somewhat more
``mobile'' and, in fact, the selection rule of\ Ref.~\cite\DG\ came
precisely from the fact that the two modes of $\l$ had to sit on the same
vertex; indeed, if $z=w$, Eq.~\eqref{QT} yields identically zero.

Now consider the world--sheet instanton correction to the \ysing\
coupling.
It will necessarily contain
 $$
   [\, \l\,\bar\l \,]_{(z_1)}\> [\, \l\,\bar\l \,]_{(z_2)}\>
     [\, \l\,\bar\l \,]_{(z_3)}~.
 $$
When substituting the zero-modes of the $\l$'s and $\bar\l$'s, at least
one
$\l$ and at least one $\bar\l$ has to be repeated and the Berezin
integration yields zero. This fate may be be evaded only if we contract a
$\l^\a(z)$ with a $\l_\b({w})$ (from different vertices). This can be
done in several different ways, each bringing down a $\d^\a_\b(X) /(z-w)$
factor. Had we not lowered the indices on $\l^{\bar\x}$ and
$\l^{\bar\eta}$, there would be factors of $g^{\a\ba\b}$ instead of
$\d^\a_\b$.

Consider now the world--sheet instanton correction to the mixed couplings.
We can readily evaluate both the $\VEV{V^I_{({-}1)}\,
V^{\bar J}_{({-}1)}\,V^1_{(0)}}$ and the
$\VEV{V^0_{({-}1)}\,V^{\overline{0}}_{({-}1)}\,V^1_{(0)}}$ coupling.
In the latter one, the
$\l$'s and the $\bar\l$'s are distributed as follows
 $$
   [\, \l\,\l \,]_{(z_1)}\> [\, \bar\l\,\bar\l \,]_{(z_2)}\>
     [\, \l\,\bar\l \,]_{(z_3)}~.
 $$
It suffers from the same abundance in $\l$'s and $\bar\l$'s as the
instanton correction to the \ysing\ one, so it must vanish,
unless we contract a $\l$ from the first vertex with any of the
$\bar\l$'s.
(Contracting the $\l$ in the third, $E_6$--\sing\ vertex does not help
since the remaining two $\l$'s are both $\l^3$ and yield zero if they
occur
at the same vertex.)

Let us now consider the various contributions to the correlation
functions. The {\bf10.$\overline{\bf10}$.1} part of the \ymix\ coupling is
proportional to
 $$
   [\, \l^I\,\bar\l \,]_{(z_1)}\> [\, \l^{\bar J}\,\l \,]_{(z_2)}\>
     [\, \l\,\bar\l \,]_{(z_3)}~ =
   {\d^{I\bar J} \over (z_1 - z_2)}
   [\, \bar\l \,]_{(z_1)}\> [\, \l \,]_{(z_2)}\>
     [\, \l\,\bar\l \,]_{(z_3)}~.
 $$
We may choose the 0-picture for either of the three vertices,
so there are really three such couplings. A straightforward
substitution gives
 $$
   \eqalign{
\VEV{ V^I_{({-}1)}\, V^{\bar J}_{({-}1)}\, V^1_{(0)} }
 &= \int \rd^6z~{ -1 \over (\ba{z}_1-\ba{z}_2)}
                { +\d^{I\bar J} \over (z_1-z_2) }
                h_{\ba\m}{}^\a\, b_{\b\ba\n}\,
                 \O^{\ba\r\,\ba\s\,\ba\t}\, s_{\ba\r}{}^\g{}_\d       \cr
 &  \hskip5mm \times
    \int \rd^4\ba{\a}~\Big[\, - \ps^{\ba\m}_{(1)}\, \ps^{\ba\n}_{(2)}\,
                       \ps_{\ba\s}^{(3)}\, \ps_{\ba\t}^{(3)} \,\Big]
\cr
 &  \hskip10mm \times
    \int \rd^4\a~\Big[\, \l^\b_{(2)}\, \l^\d_{(3)}\,
                          \l_\a^{(1)}\, \l_\g^{(3)}    \,\Big]~,
\cr
            }
 $$
where the signs come from permuting the various fermions. The
parenthetical
indices remind us from which vertex the indexed field came and, hence, on
which $z_i$ it depends. Thus we have:
 $$
   \eqalign{
\VEV{ V^I_{({-}1)}\, V^{\bar J}_{({-}1)}\, V^1_{(0)} }
 &=  2 \d^{I\bar J}
    \int {\rd^2z_2 \over |cz_2+d|^4}~ b_{3\bar3}(X(z_2))    \cropen{8pt}
 &  \hskip-80pt    \times~
     \int {\rd^2z_1\, \rd^2z_3 \over |cz_1+d|^4 |cz_3+d|^4}~
      \bigg(\, { cz_1+d \over cz_3+d } \,\bigg)
       \bigg(\, { z_2-z_3 \over z_1-z_2 } \,\bigg)
 h_{\bar3}{}^{[1}(X(z_1))\,s_{\bar3}{}^{2]}{}_3(X(z_3)) ~,   \cr}
\eqlabel{Whew!}
 $$
where
$h^{[1}\, s^{2]} \define h^1 s^2 - h^2 s^1~.$ The other two expressions,
 $$
\VEV{V^I_{({-}1)}\,V^{\bar J}_{(0)}\,V^1_{({-}1)}}\qquad \hbox{and} \qquad
\VEV{V^I_{(0)}\,V^{\bar J}_{({-}1)}\,V^1_{({-}1)}}~,$$
yield the same result as in Eq.~\eqref{Whew!}, as do the mixed cubic
Yukawa
couplings with all the other $SO(10)\subset E_6$ components of the \gen's
and
\agen's.

Consider now the ${\bf1}^3$ coupling:
 $$
   \eqalign{
\VEV{ V^1_{({-}1)}\, V^1_{({-}1)}\, V^1_{(0)} }
 &= \int \rd^6z~{ 1 \over (\ba{z}_2-\ba{z}_1)}~
          s_{\ba\m}{}^\a{}_\b~ s_{\ba\n}{}^\g{}_\d~
                 \O^{\ba\r\,\ba\s\,\ba\t}~ s_{\ba\r}{}^\h{}_\k       \cr
 &  \hskip5mm \times
    \int \rd^4\ba{\a}~\Big[\, + \ps^{\ba\m}_{(1)}\, \ps^{\ba\n}_{(2)}\,
                       \ps_{\ba\s}^{(3)}\, \ps_{\ba\t}^{(3)} \,\Big]   \cr
 &  \hskip10mm \times
    \int \rd^4\a~\Big[\, -\l^\b_{(1)}\, \l^\d_{(2)}\, \l^\k_{(3)}\,
                           \l_\a^{(1)}\, \l_\g^{(2)}\, \l_\h^{(3)}
                                                       \,\Big]^!~.     \cr
            }
 $$
Since the three $\l_{(i)}$'s are already on different vertices,
we can contract any one of the $\l_{(i)}$'s with any of the
$\l^{(i)}$'s from either of the other two vertices. (Contracting $\l$'s
from
the same vertex would yield a pole of first order, but such
contributions do not arise as the $s_{\ba\m}{}^\a{}_\b$'s are traceless.)
The
remaining six contributions are of the form of
 $$
  -\VEV{ \l^\b_{(1)}\, \l^{(2)}_\g }
    \int \rd^4\a~
     \l^\d_{(2)}\, \l^\k_{(3)}\, \l_\d^{(1)}\, \l_\h^{(3)}~,
 $$
which evaluates to
 $$
   -{ \displaystyle
        \d^\b_\g\, \d^\d_3\, \d^\k_3\,
                (\d_\a^1\, \d_\h^2 - \d_\a^2\, \d_\h^1)
       \over
       (\ba{c}\ba{z}_1+\ba{d})^2\> (\ba{c}\ba{z}_2+\ba{d})^2\>
                               (\ba{c}\ba{z}_3+\ba{d})^2 }
      \Big(\, { z_2-z_3 \over z_1-z_2 } \,\Big)
       \Big(\, { cz_1+d \over cz_3+d } \,\Big)~.
 $$
Collecting all contributions, we get
the simple answer
 $$
   \eqalign{
\VEV{ V^1_{({-}1)}\, V^1_{({-}1)}\, V^1_{(0)} } &=
    \int { \rd^2z_1 \over |cz_1+d|^4 }\,
          { \rd^2z_2 \over |cz_2+d|^4 }\,
           { \rd^2z_3 \over |cz_3+d|^4 }\>
            \bigg(\, { cz_1+d \over cz_3+d } \,\bigg)
             \bigg(\, { z_2-z_3 \over z_1-z_2 } \,\bigg)
\cropen{8pt}
 &  \hskip25mm \times
              s_{\bar3}{}^{[1|}{}_\a(X(z_1))~
               s_{\bar3}{}^\a{}_3(X(z_2))~
                s_{\bar3}{}^{|2]}{}_3(X(z_3))~.                  \cr
            }
\eqlabel{Whew.}
 $$

Compare now the results in Eq.~\eqref{Whew!} and
\eqref{Whew.}.
On inverting Eq.~\eqref{inst.eq},
 $$
    z_i~ = - { d \z_i - b \over c \z_i - a }~,
 $$
we see that
 $$
   \bigg(\, { cz_1+d \over cz_3+d } \,\bigg)
    \bigg(\, { z_2-z_3 \over z_1-z_2 } \,\bigg)~ =
   \bigg(\, { \z_2 - \z_3 \over
              \z_1 - \z_2 } \,\bigg)~,
 $$
so that, for example,
 $$
\VEV{ V^1_{({-}1)}\, V^1_{({-}1)}\, V^1_{(0)} } =
    \int \rd^2\z_1\, \rd^2\z_2\, \rd^2\z_3~
      \bigg(\, { \z_2 - \z_3 \over
              \z_1 - \z_2 } \,\bigg) \
              \End{s}{[\x|}{\a}(\z_1)~
               \End{s}{\a}{\z}(\z_2)~
                \End{s}{|\y]}{\z}(\z_3)~.    $$
Therefore, up to the overall factor, all three
results~\eqref{Whew!} and \eqref{Whew.}, and all other
components of the {\bf27.$\overline{\bf27}$.1} and ${\bf1}^3$ couplings
are of
the form
 $$
   \int \rd^2\z_1\, \rd^2\z_2\, \rd^2\z_3\,
      \bigg(\, { \z_2 - \z_3 \over
            \z_1 - \z_2 } \,\bigg) \
   A(\z_1)\, B(\z_2)\, C(\z_3)~.            \eqlabel{derivative}
 $$
We turn now to a discussion of the form of these expressions.
\subsection{The form of the Yukawa couplings}
The expression we obtain for the ${\bf 1}^3$ coupling is proportional to a
sum of terms of the form
 $$
\k(a,b,c)\define
\half\ve_{ij}
\int d^2\z_1d^2\z_2d^2\z_3\;\left({\z_2-\z_3\over \z_1-\z_2}\right)
\End{a}{i}{\m}(\z_1) \End{b}{\m}{\z}(\z_2) \End{c}{j}{\z}(\z_3)
\eqlabel{singlets1}
 $$
where the indices $i$ and $j$ run over the transverse coordinates $\x$ and
$\y$.
The $\End{b}{\m}{\z}$ here should not be confused with the (1,1)-form
$b_{\m\bar\n}$.

We make use also of the full form of Cauchy's Theorem:
 $$
\int_{\partial B}{d\z\over\z}\vph=2\p i\vph(0)-
\int_B {d\z\over\z}\delb\vph  \eqlabel{Cauchy}
 $$
which is a consequence of the identity
 $$
\delb\left({d\z\over\z}\right) = -\p\d^{(2)}(\z)\,d\z d\bar\z~,
 $$
where $\d^{(2)}(\z)$ means $\d(\real\,\z)\d(\imag\,\z)$, and Stoke's
Theorem. Cauchy's Theorem is of course most often applied to holomorphic
functions in which case the integral on the right hand side of
\eqref{Cauchy} vanishes. For our present purpose we wish to apply
\eqref{Cauchy} on the world--sheet\footnote{$^2$}{or rather to its image
in
the \cym, the rational curve, which is also an $S^2=\cp1$}, $\S$,
so we take $B$ to be a disk of radius $R$ and let $R\to\infty$. If we
assume also that $\vph$ approaches a finite limit at $\infty$ then the
boundary contribution becomes $2\p i\vph(\infty)$ and hence
 $$
\int_\S {d\z\over\z}\delb\vph=2\p i(\vph(0) - \vph(\infty))~.
 $$
Proceeding in this manner we find that the integral in \eqref{singlets1}
is
proportional to
 $$
\ve_{ij}\int_{\S_1\times\S_2}(a\times b)^{i}(\z_1)\,
c^{j}(\z_2)\,(\z_1-\z_2)\eqlabel{singlets2}$$
with
 $$
(a\times b)^j=a^j \b - \a^j{}_k b^k~~~,~~~
a^i=\End{a}{i}{\z}d\z d\bar\z~~~,~~~
b^i=\End{b}{i}{\z}d\z d\bar\z~~~,~~~
c^i=\End{c}{i}{\z}d\z d\bar\z~.
 $$

Note that the above expression~\eqref{singlets2}\ is non-zero because
the surface term in~\eqref{Cauchy}\ does not vanish. This
phenomenon is related to the existence of unphysical particles whose
contribution to the path integral vanishes as it takes the form of
a total derivative. However, when the surface term, as above, does not
vanish
they become physical. In order to remedy this and guarantee unitarity
one needs to take the contribution from contact terms into account in
order
to cancel the surface terms~~\Ref\GS{M.~Green and N.~Seiberg, \npb{299}
(1988) 559.}. Let us now turn to this in more detail.

The Yukawa coupling that we have calculated has the unattractive
feature that  it seems not to be well-defined on BRST cohomology
classes, that is, it seems to depend on the precise representatives
chosen  for the cohomology classes in ${\rm H}^1({\rm End}{\cal T})$.
On the  other hand, it is also given by a sum of $\delta$-function
contributions where two of the three vertex operators coincide. This
latter property is the key to ``curing" the above ambiguity.

Recall that, in general, the correlation functions of vertex operators
are  ambiguous up to the addition of contact terms when vertex
operators  collide. Generally, one can fix these contact terms by
demanding  that the Ward identities of the theory be preserved. In the
above  calculation, we blithely assumed that the contact terms
vanished. We  found, unfortunately, that the resulting amplitude was
not  well-defined on BRST cohomology classes.
Clearly then, {\it zero} is the wrong choice for the contact terms.
The  correct choice -- the one which renders the amplitude
well-defined on  BRST cohomology classes -- is to add to the amplitude
a  contact contribution which precisely cancels (6.6).

\noindent
There are several things to be noted about the proposed contact terms.
\item{$\bullet$~~} The ambiguous amplitude (6.6) had no ``bulk"
contribution. It
was  given entirely by a contact interaction. It is only in this
felicitous  circumstance that one can cancel it by a choice of contact
terms.
\item{$\bullet$~~} The contact term, for a given representative
$a^\mu{}_\nu$, depends on the restriction of $a^i{}_j$ and
$a^\zeta{}_\zeta$  to the curve. For any given curve, there always
exists a  choice of representative such that these vanish (and hence
so does  the contact term) when restricted to that curve.
\item{$\bullet$~~} The mass term of Sections 4 and 5 depends on the
restriction of
$a^i{}_\zeta$ to the curve. Indeed, it depends only on the cohomology
class  of $a^i{}_\zeta$. It is unambiguously-defined, and receives no
contribution from the contact term discussed here.

\noindent
It would be desirable to be able to write a universal expression
for the  contact term which reduces to the right thing when restricted
to any given curve, but we have not been able to do so.

Finally, consider the \ymix\ coupling. The expression \eqref{Whew!},
reduces to (omitting the $E_6$ indices)
$$\k(h,b,s) = {1\over 2}\ve_{ij} \int \rd^6\z
          \left( {\z_2-\z_3\over \z_1-\z_2} \right) \
           h_{\bar\z}{}^{i}(\z_1)\,
           b_{\z\bar\z}(\z_2)\,
           \End{s}{j}{\z}(\z_3) ~.$$
We may set
$$ b = b_{\z\bar\z} d\z d\bar\z ~,~
   s^j =  \End{s}{j}{\z} d\z d\bar\z ~,~ {\rm and}~
   h_{\bar\z}{}^{i} d\bar\z= \delb\chi^i$$
and integrate by parts to find
$$ \k(h,b,s) = {1\over 2}\ve_{ij} \int_{\S_1\times\S_2}
              (\z_1-\z_2) \,\chi^{i}(\z_1) \, b(\z_1)\,
               s^{j}(\z_2) ~.$$
As for ${\bf 1}^3$, the surface terms is canceled by taking the contact
terms into account. Thus, we find that to first order in the
instanton expansion both ${\bf 1}^3$ and \ymix\ vanish. This is in
perfect agreement with the general argument of Section~5.3--that all
\Y\ couplings involving \sing's vanish to all orders.
Though it would clearly be preferable to give a clearer account of the
r\^ole of the contact terms.

\newpage
\section{conc}{Conclusion}
\vskip-20pt

\subsection{Summary and discussion}
Complementing the extensive and rather complete understanding of the
massless
\gen's and \agen's and their Yukawa couplings in a \cy\ compactification,
we
have examined the situation with the \sing's.

Some striking differences emerge already at the classical level: foremost,
that the number of \sing's is {\it not} constant when the complex
structure is
varied, but jumps at special $R$-symmetric subregions; see Section~2.2. By
mirror symmetry, the same is expected with respect to variations of the
K\"ahler class. Unfortunately, this means that the total space of
parameters
associated to the \gen's, \agen's and \sing's is not as uniform as the
well
studied \gen+\agen's subspace. Nevertheless, for the generic part of this
total moduli space where the number of \sing's is constant, the \ysing\
and
\ymix\ couplings can be shown to vanish among the \sing's which have a
deformation-theoretic representation. Thus, at least these \sing\
parameters
represent integrable deformations.

We then turn to the first order instanton corrections, which depend on the
families of lines (genus-0 curves) of degree 1 in the \cy\ space $\cal M$.
A generic,
$\ca{O}(-1)\oplus\ca{O}(-1)$ line is found to produce a nonvanishing entry
in
the mass matrix for the \sing's. However, as the complex structure of
$\cal M$ is varied, these lines move about and their contributions were
seen to
be finite (though not zero) in a simple case where two or more lines
in  $\cal M$ coalesce into a single line. However, we present general
arguments that the sum over all instantons (at any given order) vanishes.

The first order instanton corrections to the \ysing\ and \ymix\ Yukawa
couplings were all found to be of the same generic form. Moreover, these
corrections can (and should) be eliminated through a judicious inclusion
of
contact terms. This has the virtue of preserving not only unitarity, but
also
the vanishing of (some of the) \ysing\ and \ymix\ Yukawa couplings and so
also
the integrability of the \sing-deformations found at the classical level.

Finally, following~\cite{\SW}, in Section~5.3 we give a general
argument that all terms in the superpotential involving the \sing's
vanish to all orders. There we also discuss the possible limitations of
this argument.

\subsection{Open questions}
Some of the open questions raised by the considerations of this
article are:
\bigskip
\item{$\bullet$~~}We do not consider the expressions that we have
found for the instanton
contributions to the \ysing\ and \ymix\ couplings to be satisfactory. The
reader may care to set this matter in order.
\medskip
\item{$\bullet$~~}The present discussions of the mass-matrix and the
Yukawa
couplings leaves untouched the difficult and important issue of the
kinetic
term for the \sing's. That is, we need a better understanding of the
geometry of the space of singlets, \ie the analogue of special geometry.
\medskip
\item{$\bullet$~~}It remains unclear how to sum up explicitly the
contributions of all
the lines to the mass-matrix even for a simple model such as $\cp4[5]$.
\medskip
\item{$\bullet$~~}A technical point: the expression~~\eqref{polend}\
that we use to relate polynomial-forms $\hbox{\ssa}_M dx^M$ to elements of
$H^1(\M, \hbox{End}\>\T)$ is in reality an ansatz. It would be nice to
have a derivation of this via a residue calculation, say, as
in~\cite{\rPhilip}\ for the polynomial deformations of $H^1(\M,\T)$.
 %
 %
\vskip2.3truein
\noindent{\bf Acknowledgments}:\\
It is a pleasure to thank S.~Katz, B.~Greene, D.~Morrison,
E.~Silverstein  and E.~Witten for
instructive discussions.
 P.~B. was supported by the American-Scandinavian Foundation, the
Fulbright
Program, NSF grants PHY 8904035 and PHY 9009850, DOE
grant DE-FG02-90ER4052 and the Robert~A.~Welch Foundation. P.~B. would
also
like to thank the ITP, Santa Barbara and the Theory Division, CERN for
their
hospitality during part of this project. In the various stages of this
project,
T.H.\ was supported by the DOE grants DE-FG02-88ER-25065 and
DE-FG02-94ER-40854, and the Howard University 1993 FRSG Program.
 P.~C. was supported by NSF grants PHY 9009850 and PHY 9021984 and the
Robert~A.~Welch Foundation. E.~D. was supported by NSF grant PHY
9009850, the Robert~A.~Welch Foundation and by the Alexander von
Humboldt-Stiftung.
Thanks are also due the M.S.R.I.,
Berkeley, for hospitality to the authors while this work was begun.
The work of X.~D. was supported by a grant from The Friends of The
Institute for Advanced Study.
X.~D. would also like to thank the Escuela de F{\'\i}sica of the
University
of Costa Rica where some of the work presented here was done.
\newpage
\chapno=-1
\section{appendix}{Remarks on the Determinants and Resolution of
an Apparent Paradox}
It is instructive to inquire how the individual factors in $\m$ vary
with  $\e$.
We first choose a basis of zero modes
 $$
\z_r= \pd{\z}{a^r}~,\qquad
\ps_\r= \pd{\ps}{\a^\r}~,\qquad
\l_{\bar\s}=\pd{\l}{\a^{\bar\s}}~, \eqlabel{zmodebasis}
$$
and set
 $$
G^B_{r\bar s} = \int {d^2 z\over\n^2} g_{\z\bar\z} \z_r
        \bar\z_{\bar s}~,\qquad
G^F_{\r\bar\s} = \int {d^2 z\over\n} g(\ps_\r, \l_{\bar\s})~,
\eqlabel{metrics}$$
(the single factor of $\n$ in the second metric is due to the inclusion of
a
factor of $(h^{zz})^{\half}$ that takes account of the fact that the
fermionic zero modes are spinors).
Although this is not immediately apparent from~~\eqref{metrics}\ it is
straightforward to check that $G^B$ and $G^F$ are in fact independent
of the parameters $(a,b,c,d)$ of the instantons.
The bosonic metric, $\det(G^B)$ has a limit as $\e\to 0$ that is finite
and non
zero. The behaviour of the other factors is however more interesting.
By taking
the metric $g$ that appears in \eqref{metrics} to be the restriction on
the
Fubini--Study metric to the \cym, it is straightforward to compute the
dependence of $G^F_{\r\bar\s}$ on $\e$.
Consider for example the case $\k=0$ for the lines~$l_\e$ of Section~5.2.
We find that for such a line
 $$
\det G^F \approx  |\e|^{-4}~.
 $$
(The symbol $\approx$ means asymptotic equality up to multiplication by a
constant.) Since we have argued that $\m$ is constant it must therefore be
the
case that
 $$
{\hbox{Det}'(\bar\partial^\dagger_{\ca{T}\otimes\ca{S}}
            \bar\partial^{\vphantom\dagger}_{\ca{T}\otimes\ca{S}})  \over
  \hbox{Det}'(\bar\partial^\dagger_\ca{T}
            \bar\partial^{\vphantom\dagger}_\ca{T}) }\approx |\e|^{-4}~.
\eqlabel{detprime} $$
{}From \eqref{detprime} we see that the basis \eqref{instantini}
that we have been using for the
fermionic zero modes is singular as $\e\to 0$.  It is curious that this is
the best choice of basis  (since for this basis $\m$ is constant).
Consider
then the following basis, which might seem a better choice ($\ps^{\bar\z}$
and
$\l^{\z}$ are as before)
 $$\eqalign{
\ps_{\bar\xi} &= - {\ba{\a}\bar\epsilon^2 \over \bar c \bar z+\bar d}
\cropen{3pt}
\ps_{\bar\eta} &= {\ba{\a} \bar a\over \bar c(\bar c \bar z+\bar d)} +
{\ba{\b} \over \bar c \bar z+\bar d}\cr}
\hskip30pt
\eqalign{
\l_{\xi} &= - {\a\epsilon^2 \over c z + d} \cropen{3pt}
\l_{\eta} &= {\a a\over c(c z + d)} +
{\b \over c z + d}\cr}    \eqlabel{movingbasis}   $$
Now  given the relation
 $$\eqalign{
\x - {\z\eta\over {\e^2}} &\rightarrow \x_0 \cr
\eta                    &\rightarrow \eta_0 \cropen{3pt}
\z                      &\rightarrow \z_0~.\cr}\eqlabel{coordchange}$$
between the coordinates
$(\xi,\eta)$, for the line $l_{+\e}$ of type \splittingtype{-1}{-1}, and
$(\xi_0,\eta_0)$, for the line of type \splittingtype{0}{-2}, we find,
to  leading
order,
 $$\eqalign{
{\partial \over \partial\xi_0}&\sim {\partial \over \partial\xi}~,\cr
{\partial \over \partial\eta_0}&\sim
{\zeta \over \epsilon^2}{\partial \over \partial\xi} +
{\partial \over \partial\eta}~.\cr}$$
It follows that
 $$\eqalign{
\ps_{\bar\xi_0} &\sim \ps_{\bar \xi} \sim
- {\ba{\a}\epsilon^2 \over \bar c \bar z+\bar d}\to 0~,\cr
\ps_{\bar \eta_0}&\sim {\zeta \over \epsilon^2}\ps_{\bar \xi} +\ps_{\bar
\eta}
\to {\ba{\a} \over \bar c(\bar c \bar z+\bar d)^2} +
{\ba{\b} \over \bar c \bar z+\bar d}~.\cr}$$
In this basis $\det G^F$ has a finite and nonzero limit as $\e\to 0$.  So
now
$\m\approx |\e|^{-4}$.  However, the reader may easily repeat the steps in
the
computation of the mass matrix for the new basis.  The only change is the
replacement $m\to |\e|^4 m$ so the product $m\m$ is unchanged as it must
be.

We learn that the pole term in the mass matrix owes its existence to an
eigenvalue of the bosonic operator which vanishes as $\e\to 0$. This in
turn
we understand from the fact that $h^0(\splittingtype{-1}{-1}) = 0$ but
$h^0(\splittingtype{0}{-2}) = 1$ from which we see that the bosonic
operator
acquires a new zero mode precisely at $\e=0$. In fact we may understand in
the same way why an \splittingtype{1}{-3}\ line does not give rise to a
pole.
For an \splittingtype{1}{-3}\ line there are two new zero modes for the
bosonic operator but also two new zero modes for the fermionic operator.
These cancel so $\m$ has no pole. The zero mode calculation gives a factor
that tends to zero. Thus an \splittingtype{1}{-3}\ line gives a vanishing
contribution to the mass matrix as anticipated in \chapref{mass}.2.
It is interesting that the order of the degeneration of the instanton
metric is related to the number of zero modes of the operators
appearing in~~\eqref{detprime}.

We can now dispose rather easily of an apparent paradox that appears when
one
tries to calculate the mass matrix directly for an ${\cal O}(0)\oplus{\cal
O}(-2)$ line.  In this case, one would choose fermion zero modes
 $$\eqalign{
\ps_{\bar\xi_0} &= 0~,\cr
\ps_{\bar \eta_0}&={\ba{\a} \over \bar c(\bar c \bar z+\bar d)^2} +
{\ba{\b} \over \bar c \bar z+\bar d}\cr}
\hskip30pt
\eqalign{
\l_{\xi_0} &= 0~,\cr
\l_{\eta_0}&={\a \over  c( c z+ d)^2} +
{\b \over c z+ d}~.\cr}\eqlabel{lzeromodes}$$
and one finds by repeating the steps of \SS\chapref{mass}.1
that the contribution of the zero modes vanishes.
So it is tempting to conclude that the contribution
of such a line to the mass matrix is zero.  This, however, is a trap for
the
unwary.  If instead of setting $\e=0$ from the outset we let $\e\to 0$ and
employ the basis \eqref{movingbasis} then, as we have seen above, the zero
mode
calculation gives a contribution
 $$
m^0\approx {|\e|^4\over\e^2} = \bar\e^2~,
 $$
which indeed tends to zero with $\e$.  The contribution of the nonzero
modes,
however, is $1/|\e|^4$ so the asymptotic behaviour is in fact $1/\e^2$.

It is instructive also to consider now the ${\bf \overline{27}}^3$
coupling
from the
same perspective. The same apparent paradox arises if one tries to
calculate
the
contribution of an \splittingtype{0}{-2}\ line by integrating directly
over the
double line. The zero modes \eqref{lzeromodes} are such that the zero mode
integral
vanishes. However we know that the contribution of a double line to the
${\bf \overline{27}}^3$ coupling is not zero. It is twice the contribution
of a single line; and all the single lines contribute an equal and
non-zero
amount. The resolution is as before: if we work with the moving basis
\eqref{movingbasis} then, as is easily seen, the zero mode integral is
proportional to $|\e|^4$ while $\m\approx |\e|^{-4}$, due as before to the
fact that one of the eigenvalues of the bosonic operator tends to zero as
$|\e|^4$. In fact by turning these considerations around we could have
seen
sooner that the bosonic operator must have an eigenvalue that vanishes as
$|\e|^4$ and from this the $1/\e^2$ pole in the mass term follows easily
and
inevitably.
\newpage
\immediate\closeout\referencewrite
                     \referenceopenfalse
                      \line{\bf\hfil References\hfil}\bigskip
                       \parindent=0pt\input referenc.texauxil
\bye